\documentclass[conf]{new-aiaa}
\usepackage[utf8]{inputenc}

\usepackage{graphicx}
\usepackage{tikz}
\usepackage{pgfplots, pgfplotstable}
\pgfplotsset{compat=1.14}
\usetikzlibrary{pgfplots.groupplots, shapes}
\usepackage{subcaption}
\usepackage[skip=1.5pt]{caption} 
\usepackage{tikzit}

\tikzstyle{Major State}=[fill=white, draw=black, shape=circle, tikzit category=State, tikzit fill=white, tikzit draw=black, tikzit shape=circle]
\tikzstyle{Minor State}=[fill=white, draw=black, shape=circle, tikzit category=State, tikzit fill=white, tikzit draw=black, tikzit shape=circle]
\tikzstyle{Annotation}=[fill=white, draw=white, shape=rectangle, tikzit category=Annotation, tikzit shape=rectangle, tikzit draw={rgb,255: red,128; green,128; blue,128}, tikzit fill=white]

\tikzstyle{Directional Egde}=[->]

\usepackage[capitalize]{cleveref}
\usepackage{amsmath}
\usepackage[version=4]{mhchem}
\usepackage{siunitx}
\usepackage{longtable,tabularx}
\usepackage{booktabs}
\usepackage{pdflscape}
\title{Analysis of Fleet Management and Network Design for On-Demand Urban Air Mobility Operations}

\author{Sheng Li\footnote{Ph.D. Candidate, Department of Aeronautics and Astronautics, AIAA Student Member, \url{lisheng@stanford.edu}}}
\affil{Stanford University, Stanford, CA, 94305, USA}
\author{Maxim Egorov\footnote{Research Scientist, Autonomy at Airbus UTM, \url{maxim.egorov@airbus-sv.com}}}
\affil{Airbus UTM, Sunnyvale, CA, 94086, USA}
\author{Mykel J. Kochenderfer\footnote{Associate Professor, Department of Aeronautics and Astronautics, AIAA Associate Fellow, \url{mykel@stanford.edu}}}
\affil{Stanford University, Stanford, CA, 94305, USA}

\begin{document}

\maketitle

\begin{abstract}
A significant challenge in estimating operational feasibility of Urban Air Mobility (UAM) missions lies in understanding how choices in design impact the performance of a complex system-of-systems. 
This work examines the ability of the UAM ecosystem and the operations within it to meet a variety of demand profiles that may emerge in the coming years. 
We perform a set of simulation driven feasibility and scalability analyses based on UAM operational models with the goal of estimating capacity and throughput for a given set of parameters that represent an operational UAM ecosystem. 
UAM ecosystem design guidelines, vehicle constraints, and effective operational policies can be drawn from our analysis. 
Results show that, while critical for enabling UAM, the performance of the UAM ecosystem is robust to variations in ground infrastructure and fleet design decisions, while being sensitive to decisions for fleet and traffic management policies. 
We show that so long as the ecosystem design parameters for ground infrastructure and fleet design fall within a sensible range, the performance of the UAM ecosystem is affected by the policies used to manage the UAM traffic. 
\end{abstract}

\section{Introduction}

\lettrine{T}{echnological} advances in electrical vertical takeoff and landing (eVTOL) aircraft have enabled a variety of new missions that could soon be feasible including cargo delivery and passenger transportation. 
Some estimates show that by the year 2035, the hourly demand for passenger carrying urban flights could be as large as 2,500 in metropolitan areas like Paris~\cite{karthik2018blueprint}. 
Passenger carrying Urban Air Mobility (UAM) operations would ultimately try to capture some fraction of this demand.
However, designing a system capable of safe and efficient UAM operations at such a scale is a major challenge.
While certain aspects of eVTOL vehicle design and requirements have become more concrete in recent years~\cite{clarke2019strategies, kadhiresan2019conceptual}, the uncertainty surrounding what the UAM ecosystem as a whole will look like still exists.

For UAM operations and its supporting systems, the majority of design decisions will be driven by the safety and business cases. 
However, it is difficult to evaluate the implications of a design choice on these two factors without examining the UAM ecosystem as a whole.
Specifically, understanding how various constraints and design choices interact with one another is critical to understanding the performance of the system. 
For example, decisions on the capacity of an in-demand vertiport, the size of a fleet used to serve a region, and the choice of policies used to manage traffic all have a crucial impact on performance metrics like vehicle throughput and demand fulfillment rate of the UAM ecosystem in a given region. 
Understanding how these decisions impact the overall performance of the system can inform a more efficient design process for UAM, leading to a safer, healthier, and more sustainable ecosystem. 

The UAM ecosystem is comprised of a number of independent but interconnected systems, including the supporting ground infrastructure like a network of vertiports, the UAM vehicle fleet, and the system used to manage UAM traffic. 
These systems, like management and coordination of UAM traffic, can be further divided into systems responsible for managing an operator's fleet and the system responsible for coordination between various stakeholders in the system, such as a UAM equivalent to UTM~\cite{prevot2016uas}.  
Relationships between these systems are difficult to evaluate and an accurate analysis generally requires a comprehensive view of all the systems functioning together, which is usually not feasible until the ecosystem is operational. 
Challenges also emerge from the on-demand nature of UAM operations.
It is expected that some operators will not follow fixed schedules, as in current commercial air transportation systems, which can result in load imbalance problems. 
Accounting for the requirements, constraints, and inter-relationships imposed by these systems is difficult, and analyzing one of the parts in isolation could lead to poor estimates of overall system performance.  





In this work, we perform a number of quantitative analyses using a macroscopic scenario simulator of UAM operations. 
We focus on the interactions between UAM operations, the underlying infrastructure, the traffic management system, and any constraints imposed on the operations to extract emergent, system-level quantitative results like throughput and demand fulfillment.
Through careful choice of design variables, we are able to formulate the UAM ecosystem design problem as an optimization. 
By performing a variety of analyses, we are able to both identify the design variable regions that lead to solutions from which feasible UAM operations emerge, and to quantify the relationships between critical design variables like vertiport capacity and operational fleet sizes. 
Lastly, we investigate the sensitivity of the critical design variables on system performance, and demonstrate how careful choice of fleet and traffic management policies can compensate for sub-optimal design decisions related to vertiport capacity and fleet size. 
We conclude by drawing UAM ecosystem design guidelines from our analysis.
\section{Related Work}

The design of requirements and constraints for UAM systems has become a widely explored topic in recent years. 
\citeauthor{kohlman2018system}~\cite{kohlman2018system} describe a system-level model of a UAM network and explored energy-related constraints such as battery life of eVTOL aircraft as well as the number of charging stations that are part of the ground infrastructure. 
\citeauthor{vascik2018scaling}~\cite{vascik2018scaling} separately discuss scaling constraints for UAM operations from air traffic control, ground infrastructure, and noise, with additional highlights on the additional congitive loads for air traffic control that come from UAM~\cite{lundberg2018cognitive}.
\citeauthor{mueller2017enabling}~\cite{mueller2017enabling} provide a conceptual description of the methodologies for enabling the integration of on-demand operations with the existing commercial transportation airspace.
While these works provide a glimpse into requirements for UAM operations, they generally lack analysis of their feasibility in a broader UAM ecosystem.

The topic of traffic management in UAM has also gained traction recently.
New approaches for traffic management have been proposed for terminal area scheduling that account for UAM constraints~\cite{kleinbekman2020rolling}, and for managing dense traffic flows in unstructured airspace~\cite{egorov2019encounter}.
Fleet management rules and ground infrastructure constraints have been evaluated and found to result in limitations on the operation of UAM ecosystems~\cite{vascik2018scaling}. 
More generally, analyses of dense and highly utilized airspace have shown a need for strategic traffic management to minimize risk~\cite{li2019optimizing}. 
The topic of fairness and its relation to traffic management has shed light on the importance of designing proper traffic management policies in a federated system with a number of self-interested stakeholders~\cite{evans2020fairness}.

System level studies have been done for networked transportation ecosystems in the past.
The problem of fleet management and the associated behaviors in the system that emerge has been examined in on-demand autonomous car networks~\cite{zhang2016control, hyland2017taxonomy}, for autonomous drone fleets with delivery applications~\cite{lee2017optimization}, and in multi-modal transport systems~\cite{choudhury2019efficient, salazar2019intermodal}.
These approaches consider a number of system-level constraints that may be applicable to UAM operations.
However, they provide limited analysis of design system-of-system wide design decisions and their relationships in the broader transportation ecosystem. 
Certain frameworks do examine the transportation network design problem from an optimization perceptive~\cite{di2018transportation}, and consider how the impact of  constraints and design decisions impact emergent, system-wide performance characteristics ~\cite{ding2017heuristic}.
The design objectives generally do not consider characteristics of interest typical to UAM such as on-demand operations and limited constraints at network nodes, leaving feasibility and performance of an urban air transportation systems poorly understood.  
Approaches for maximizing throughput in transportation networks have been proposed as well~\cite{williams2010system, chen1999capacity}, however little work has been done on analyzing the constraints that are critical to UAM related to vertiports and the on-demand nature of the operations.
A recent collection of system-levels analysis for UAM~\cite{vascik2020systems} is perhaps the most closely related work to what is presented in this paper. 
While that work provides a comprehensive analysis of operation scaling constraints for UAM, it provides quantitative simulation results at a single vertiport level and does not evaluate performance of the UAM ecosystem as a whole. 
Overall, a system level quantitative analysis and design of the UAM ecosystem is not widely covered in existing literature, and this work aims to fill that gap.

\section{Problem Formulation}\label{sec:problem_formulation}
\begin{figure}[ht!]
    \centering
    \includegraphics[scale=0.17]{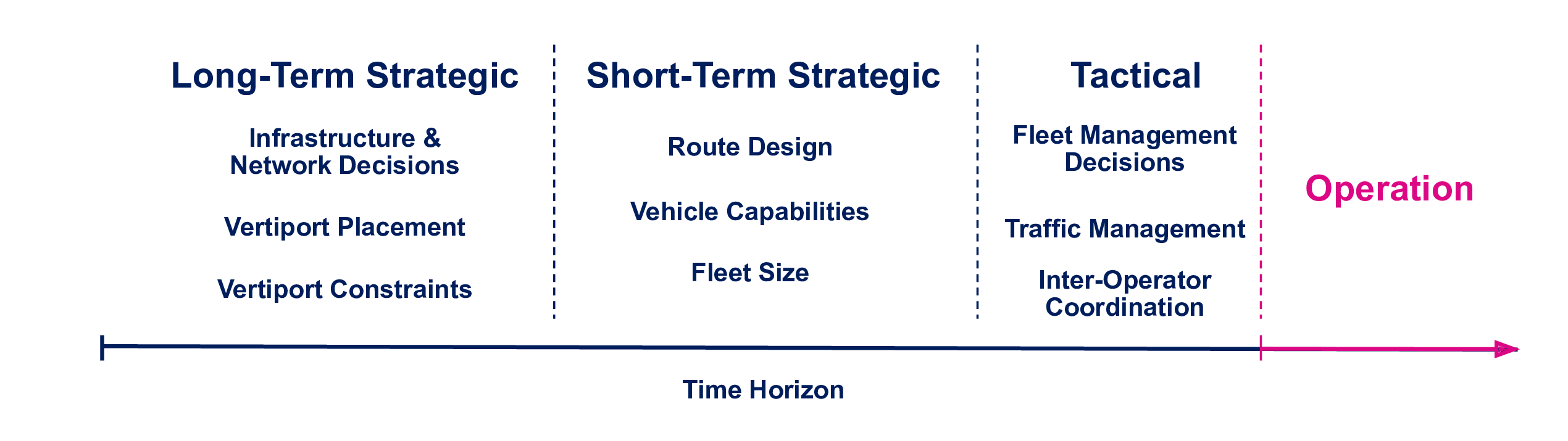}
    \caption{UAM design decision variables and the associated time-horizon for when these decisions may be made.}
    \label{fig:time_sacle}
\end{figure}

In this work, we are interested in the problem of UAM ecosystem design.
To that end, we want to evaluate the performance of the UAM ecosystem with respect to a number of design variables that correspond to critical ecosystem components.
Specifically, we want to evaluate the impact that design variables of interest have on performance metrics such as demand throughput, average passenger delay, and design variable sensitivities.  
We focus on nominal operations that are disturbance free.
Additionally, we model stochastic demand profiles, and uncertainty driven properties of the variables of interest such as vehicle turn-around time at vertiports. 
We leave system degradation events that could emerge from vehicle system failures or weather events for future work. 
We organize the design variables into three categories, roughly based on the time-horizon from the start of a UAM operation when these variables must generally be chosen. 
The categories are illustrated in~\cref{fig:time_sacle} and are described further below. 

We split the design variables into long-term strategic, short-term strategic, and tactical categories. 
The long-term strategic variables are those that take on the order of years to modify and are most commonly related to the infrastructure of a UAM ecosystem.
The short-term strategic variables are attributed to the UAM fleet and the route design in a network, requiring months to change.
The tactical decisions are related to the fleet and traffic management components and may be made on the order of days or even hours. 
The above categorization of the variables helps shed light on the impact of long-term, short-term, and tactical decision on the performance of the UAM ecosystem, with the goal of understanding at which point in the design cycle certain decisions must be made, and at which point they begin to impact decisions downstream.
We describe the design variables that are appropriate for each category in the sections below, and summarize them in~\cref{tab:design-vars}.
In the following sub-sections, we describe the design variables analyzed in this work and the time categories they fall into, as well as their associated modeling assumptions. 

\begin{table}[]
\caption{UAM ecosystem design variables considered in this work and whether they are frozen or varied in our analyses.}
\label{tab:design-vars}
\centering
\small
    \begin{tabular}{lcc}
        \toprule
        Design Variable  & Variable Type & Variable Frozen  \\ \midrule
        Vertiport Capacity & Long-Term Strategic & Varied \\
        Vertiport Locations & Long-Term Strategic & Frozen \\
        Vertiport Ops Constraints & Long-Term Strategic & Frozen \\
        Fleet Size & Short-Term Strategic & Varied \\
        UAM Route Network & Short-Term Strategic & Frozen \\
        Fleet Management Policy & Tactical & Varied \\
        \bottomrule
    \end{tabular}
\end{table}



\subsection{Long Term Strategic Design Variables}
We consider a design variable to be long-term strategic when it would take on the order of years to modify it substantially. 
While all of of the variables in the long-term strategic category considered in this work are related to the UAM ecosystem infrastructure, such as the design decisions related to vertiports, a number of other critical design variables exist in this category.
For example command and control (C2) communication links are a critical component of the UAM ecosystem that may require years of design and approval time due to it's safety critical nature and the complexity of the problem~\cite{greenfeld2019concept}. 
However, these types of design decision are typically difficult to put into variable form, and we leave their analysis and how they fit into the larger UAM ecosystem as part of future work. 

\subsubsection{Vertiport Capacity}
UAM vehicles need vertiports to land, recharge or refuel, await passengers and at times perform maintenance. 
The total capacity of vertiports in a UAM network determines the maximum fleet size that can serve that network which ultimately determines the maximum demand rate the network can serve. 
Furthermore, we introduce a variable called normalized vertiport capacity, which quantifies the relationship between the total capacity of the vertiports in a UAM network and the total fleet size used to service it.
Given the total size of a fleet serving a UAM network $f$, and the total capacity of all the vertiports that comprise the network $c$, the normalized vertiport capacity takes the following form:
\begin{equation}
    c_{n} = \frac{c}{f}.    
\end{equation}
This metric can capture a simple measure of the available space in a UAM network for a given fleet.  
For example, a network with $c_{norm}=1$ has net capacity to exactly match the fleet serving it. 
Note that highly imbalanced demand profiles that favor specific vertiports, can still create demand/capacity imbalances in UAM networks with $c_{norm}=1$.
When $c_{norm} > 1$, the UAM network has capacity that is greater than the size of the fleet using it.
On the other hand $c_{norm} < 1$ implies that a UAM network does not have sufficient capacity for the fleet serving it, and may require overflow usage of other infrastructure that may not be part of the UAM network. 
In this work, we focus on capacity to accommodate aircraft that are either idle or are being recharged or under maintenance. 
Additional variation on capacity constraints related to landing and departure fixes around a vertiport could also be considered, but are left as future work here.

\subsubsection{Vertiport Geographical Locations}
Deciding geographical locations for vertiports is a challenging problem~\cite{goyal2018urban}. 
The location choice of vertiports must typically consider trade-offs between passenger demand and noise concerns in addition to other critical factors, and can be typically parameterized by a latitude and longitude value.
Ultimately, the geographical locations of vertiports determine demand rates and aircraft travel time which are critical operational factors within the UAM ecosystem. Due to their cost and complexity, it may take years to build or relocate a vertiport. 
The constraints brought by vertiport geographical locations are, thus, considered long term.

\subsubsection{Vertiport Operation Constraints}

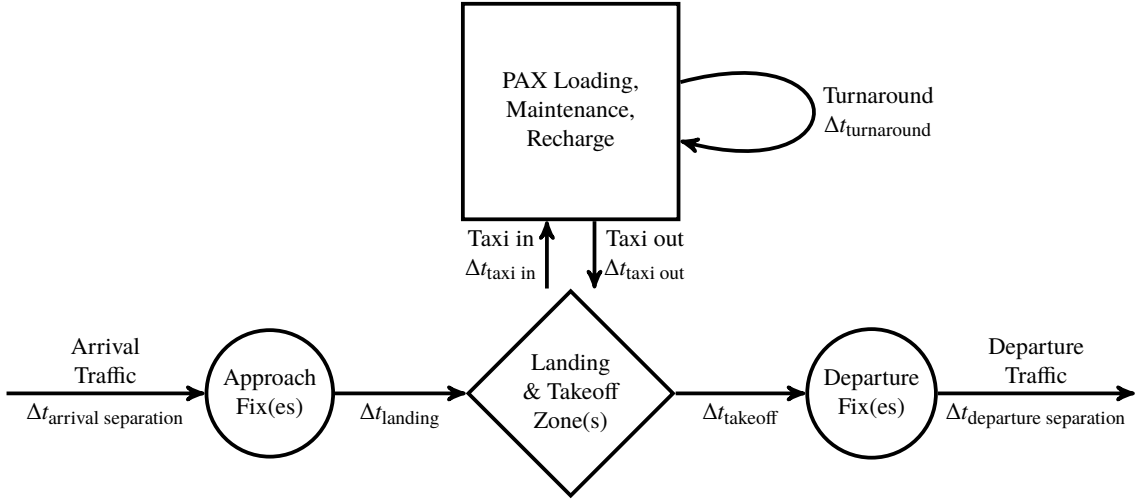
\begin{figure}[h!]
    \centering
    \begin{tikzpicture}[>=stealth', 
square/.style={regular polygon,regular polygon sides=4}]
\small
\node[diamond,
    draw=black,
    line width=0.5mm,
    align=center] (landing) at (0, 0) {Landing \\ \& Takeoff \\ Zone(s)};
    
\node[circle,
    draw=black,
    line width=0.5mm,
    align=center] (approach) at (-4, 0) {Approach \\ Fix(es)};

\node[circle,
    draw=black,
    line width=0.5mm,
    align=center] (departure) at (4, 0) {Departure \\ Fix(es)};
    
\node[square,
    draw=black,
    line width=0.5mm,
    align=center] (garage) at (0, 3.75) {PAX Loading, \\ Maintenance, \\ Recharge};

\draw[->, line width=0.5mm] (approach) to node [below, align=center] {$\Delta t_\text{landing}$} (landing); 
\draw[->, line width=0.5mm] (landing) to node [below, align=center] {$\Delta t_\text{takeoff}$} (departure); 
\draw[->, line width=0.5mm] (-7.5, 0) to node [above, align=center] {Arrival \\ Traffic} node [below, align=center] {$\Delta t_\text{arrival separation}$} (approach); 
\draw[->, line width=0.5mm] (departure) to node [above, align=center] {Departure \\ Traffic} node [below, align=center] {$\Delta t_\text{departure separation}$}  (7.5, 0); 
\draw[->, line width=0.5mm, transform canvas={xshift=-1em}] (landing) to node [left, align=center] {Taxi in \\ $\Delta t_\text{taxi in}$} (garage); 
\draw[->, line width=0.5mm, transform canvas={xshift=1em}] (garage) to node [right, align=center] {Taxi out \\ $\Delta t_\text{taxi out}$} (landing); 
\draw[->, line width=0.5mm] (garage) edge [loop right] node [align=center] {Turnaround \\ $\Delta t_\text{turnaround}$} (garage);

\end{tikzpicture}
    \caption{Vertiport abstract structure and operation flow.}
    \label{fig:vertiport_operation}
\end{figure}

Vertiport operations can be a complex process with a number off critical factors that must be considered. 
In this work, we focus on constraints related to arrival and departure procedures, in and out of vertiports, and constraints associated with the design decisions on maintenance, recharging or battery-swaps, and passenger loading. 
Because these constraints are closely tied to the vertiport design, they are considered long term. 
\cref{fig:vertiport_operation} shows the notional model used to represent a vertiport in this work. 
It is based on commonly found models in literature~\cite{vascik2018scaling}.
In the model, a vertiport consists of landing and takeoff zones and a location for maintenance and passenger loading.
For vertiports where maintenance and passenger loading can be performed directly or nearly directly to the landing zones, $\Delta t_{taxi}$ times may be negligible and can be set to zero.  
The airspace near a vertiport can be structured into approach fixes and departure fixes which are fixed approach and departure waypoints the aircraft must use.
Upon arrival, aircraft approach a vertiport through its approach fixes. 
Then aircraft land in the landing and takeoff zones, may taxi to another area, unload passengers, and may require maintenance procedures such as battery replenishment and safety inspections. 
Once the aircraft are ready for flight, they depart from the vertiport through the departure fixes. 
To maintain sufficient separation in the vertiport airspace, only a limited number of approach fixes and departure fixes can be implemented. 
They can limit the number of aircraft landing or taking-off. 
Metering can be applied at the fixes to control the traffic flow through a vertiport~\cite{gilbo1997optimizing}. 
We consider the times outlined in~\cref{fig:vertiport_operation} to be long-term design variables due to their strong link to the underlying vertiport infrastructure. 
Additionally, the separation standards around vertiports are often set in place by regulators and are enforced the traffic management system and the participating operators~\cite{goyal2018urban}. 
They directly guide the design of approach and departure fixes, which are also considered long-term design variables .

\subsection{Short Term Strategic Design Variables}
Short-term strategic design variables can be modified on the order of months.

\subsubsection{Fleet Size}\label{sec:fleet-size}
The UAM vehicle fleet is a critical component in the UAM ecosystem. 
The fleet size, denoted by $f$, can dictate a number of critical UAM operational characteristics including the maximum demand rate $d_\text{max}$ a UAM ecosystem can serve. 
The choice of fleet management policies may also be dictated by the size of the fleet serving the UAM ecosystem.
To more directly tie this design variable to others, we derive a simple upper bound $\hat{d}_\mathrm{max}$ for the maximum demand rate a fleet of size $f$ may fulfill.
It is expected that beyond this demand rate, the passenger delay times in the UAM ecosystem will increase exponentially~\cite{ota1984airport, vascik2018scaling}. 
This rate can be represented by a ratio between the total time resource of a fleet $T_\mathrm{available}$, which represents the net time the fleet is available for UAM operations, and the average end-to-end time required to complete an operation $T_\text{operation}$ by a single vehicle in the fleet. 
An estimate of the maximum demand rate is given by

\begin{equation}\label{eqn:max_demand_est}
    d_\mathrm{max}  \leq \hat{d}_\mathrm{max} = \frac{T_\mathrm{available} / \Delta t}{T_\text{operation}} =  \frac{f }{T_\text{operation}}, 
\end{equation}
where $\Delta t$ is a unit time period and $T_\mathrm{available} = f\Delta t$. 
While we refer to $\hat{d}_\mathrm{max}$ as the maximum demand rate a UAM network and the fleet operating within it can serve, it can also be considered a maximum capacity determined by the long-term and short-term strategic design variables.
In this paper, the limiting design variables dictating~\cref{eqn:max_demand_est} are the vertiport geographic locations, vertiport capacities, and the fleet size. 
Because the maximum demand rate is estimated using strategic design variables, it can also be considered the maximum throughput an effective fleet management policy can achieve in a fixed UAM network with a fixed fleet size. 
We assume fleet size to be a short term strategic constraint since it can take on the order of months to permanently increase or reduce the size of a fleet. 

\subsubsection{UAM Route Network}
The underlying routes that UAM vehicles can use to fly between vertiports comprise the UAM route network.
The placement of a route between two vertiports can depend on a number of factors such as demand, vehicle performance capability, noise considerations, and the avoidance of no-fly zones~\cite{goyal2018urban}.
Because the UAM route network ultimately determines where and how vehicles can fly, a given route structure can drive a number of critical UAM operational characteristics including the maximum demand rate $d_{max}$ a UAM ecosystem can serve.
For a fixed set of vertiports in a UAM network, it is assumed that creation or removal of routes would be possible but may require substantial efforts and approvals that could require extensive time, making this design variable short-term strategic
In this work, we assume a route network can be modeled as a graph with vertiports represented by graph nodes and the routes that connected by graph edges. 

i\subsection{Tactical Design Variables}
Tactical design variables in the UAM ecosystem are those that can be characterized as dynamic, and can be altered on the order of days or hours.

\subsubsection{Fleet Management and Re-balancing}
\label{sec:rebalancing}

The traffic management strategies primarily considered in this work are related to fleet re-balancing.
The primary function of fleet re-balancing lies in ensuring that the vehicle fleet is distributed at appropriate nodes of a transportation network in a way that enables efficient fulfillment of demand.   
Vehicle re-balancing has been well studied in ground transportation systems and networks. \citeauthor{smith2013rebalancing}~\cite{smith2013rebalancing} propose an optimal re-balancing strategy to ensure the stability of the on-demand taxi transportation system by modeling traffic flow as fluid. \citeauthor{rossi2018routing}~\cite{rossi2018routing} model a traffic congestion problem within a network flow framework and show a computationally efficient routing and re-balancing algorithm for autonomous vehicles in on-demand operation.
While fleet management and planning has been well studied in commercial aviation~\cite{clark2017buying, belobaba2009airline, tsai2012mixed}, literature for on-demand operations like UAM is limited and fleet re-balancing has not been deeply examined at the time-scales relevant to UAM.

The on-demand nature of emerging transportation systems has transformed mobility in recent years. 
It is likely that some users of the UAM ecosystem will prefer to request operations in an on-demand way as well.
Such on-demand systems can lead to downstream problems requiring fleet re-balancing.  
Specifically, a fleet that does not follow a fixed and optimized schedule, can end up distributed through the transportation network in a way where it can not meet the demand in the system.
This can lead to poor system performance and significant inefficiencies. 
Practically, this means that in extreme cases, some vertiports may be at capacity with no room left for incoming flights; while other vertiports may be empty and have no idle aircraft to carry out an incoming operation request. 
Effective fleet re-balancing can help mitigate and, in some cases, prevent these issues in the UAM ecosystem. 

While a number of re-balancing strategies are possible, in this work, we consider the following:
\begin{itemize}
    \item \textbf{Space-driven re-balancing} works by trying to keep the fleet evenly distributed within the UAM network. This also allows congested vertiports to be naturally freed up when the strategy is applied, creating a free landing space for the departing flight at its destination, and potentially minimizing any air-holding. An example scenario where space-driven re-balancing is useful is when a large number of flights arrive at a vertiport and take up all of the landing spaces. With space-driven re-balancing, flights that landed at this vertiport will takeoff to nearby vertiports with free landing spaces even if no operation request exists for that flight.
    \item \textbf{Demand-driven re-balancing} attempts to re-balance the fleet in order to meet anticipated demand in the UAM network. An example scenario where demand-driven re-balancing is useful is when a customer requests a flight at a vertiport with no available aircraft. Instead of passively waiting for the next aircraft to arrive according to incoming demand, the strategy routes an idling aircraft, which may be empty, from a nearby vertiport to carry out the operation request.
\end{itemize}
Look-ahead adjustment can be made to the two re-balancing strategies by tuning the numeric threshold used to trigger a re-balancing operation. 
A naive strategy may include simple thresholds on when the re-balancing flights take place. In this work, we improve on this heuristic and show that it significantly improves performance. 
For example, for space-driven re-balancing, a naive threshold triggers a re-balancing flight when a vertiport is full. 
The look-ahead adjustment triggers re-balancing flights prior to the vertiport filling up completely, creating a buffer for emergent conditions. 
For demand-driven re-balancing, a naive threshold is triggering re-balancing flights when a vertiport is empty. 
The look-ahead adjustment triggers re-balancing flights when the number of idling aircraft at a vertiport drops below some threshold.
In general, we expect re-balancing flights to be empty (not carrying passengers), so it is critical that they are kept to a minimum.

\subsubsection{Operation Rule: On-demand versus Scheduling}
\label{sec:scheduling}
Operating on-demand is a natural way to answer stochastic requests for UAM missions, particularly when those requests are sparse~\cite{patterson2018proposed, kohlman2018system}. 
Its counterpart is a scheduled operation where flights depart at predetermined times~\cite{lohatepanont2004airline}.
Both strategies have to consider a number of trade-offs, and one may be more appropriate for a certain operational profile than another. 
For example, on-demand operations generally require fleet re-balancing which consumes a portion of fleet resource; while scheduled operation can lead to long delays especially when requests are sparse.

\subsection{Design as Optimization}

Using the design variables described in the previous sections, we can formulate UAM ecosystem design problem as an optimization. 
While a number of critical design choices exist in the UAM ecosystem, we chose the following in this work: the UAM fleet represented by a set of UAM vehicles $f$, the design choices of individual vertiports in the ecosystem $\mathcal{V}$, vertiport network configuration $\mathcal{N}$, and the fleet management policy $\Pi$. For a given demand model $\mathcal{D}$, we want to minimize the a cost $J$ with respect to the design variables, and subject to constraints on design variables, as well as constraints on UAM ecosystem performance such as customer satisfaction, energy and safety as given in \cref{eqn:optim_general}. 
\begin{equation}\label{eqn:optim_general}
\begin{aligned}
& \underset{f, \mathcal{V}, \mathcal{N}, \Pi}{\text{minimize}}
& & J(f, \mathcal{V}, \Pi, \mathcal{N}, \mathcal{D}) \\
& \text{subject to}
& & \text{feasible vehicle design ($f$, $\mathcal{V}$, $\mathcal{N}$)}, \\ 
& & & \text{feasible infrastructure cost ($f$, $\mathcal{V}$, $\mathcal{N}$)}, \\
& & & \text{separation } (f, \mathcal{V},\Pi, \mathcal{D}, \mathcal{N}) \geq \text{minimum safe separation requirements}.
\end{aligned}
\end{equation}

In the following sections, we consider the various properties of a cost function appropriate for such an optimization.
Typical cost profiles could include operational metrics such as system wide delay or average monetary cost of a flight.
Additionally, the monetary costs of infrastructure related to $f$, $\mathcal{V}$, and $\mathcal{N}$ could be considered. Constraints of the optimization problem include but are not limited to feasible vehicle design, feasible infrastructure cost, and minimum separation between vehicles for safety consideration.
In this work, we consider constraints for a minimum separation requirement and constraints that bound UAM vehicle range within expected bounds~\cite{silva2018vtol}. 
By ensuring that distances between vertiports fall within vehicle range bounds, we ensure that constraint is met.

\section{Simulation Based Analysis for UAM Ecosystem Design}
It is difficult to quantify the operational performance of the UAM ecosystem using a single mathematical model. 
By using simulation, we are able to model individual components of that ecosystem and evaluate its performance metrics through emergent operational properties. 
Using this approach, we are able to 1) analyze the feasibility of UAM ecosystem operations, 2) analyze the impact of demand-capacity balancing on operational performance, 3) analyze the impact of infrastructure and fleet constraints on the ecosystem, and 4) optimize the performance of the network with respect to design parameters. 

\subsection{Simulation Modeling Assumptions}
We use an event driven simulator in this work. 
The simulator models high-level operational states a vehicle might find itself in under a nominal operational profile such as takeoff, landing, and maintenance.
We represent each entity using a probabilistic model known as a Markov process, where each transition between states and the times spent in each state are governed by a probabilistic model. 
Uncertainties regarding state transitions or times of transition can be incorporated directly into the probabilistic representation used in this work. 

\begin{figure}[h!]
    \centering
    \begin{tikzpicture}[>=stealth']
	\begin{pgfonlayer}{nodelayer}
		\node [style=none, align=center] (0) at (0, 0) {Aircraft Event Cycle};
		\node [style=Major State, line width=0.5mm] (2) at (-3, 0) {Arrival};
		\node [style=Major State, line width=0.5mm] (3) at (3, 0) {Departure};
		\node [style=Minor State, line width=0.5mm] (4) at (0, 1.5) {};
		\node [style=Minor State, line width=0.5mm] (5) at (-1, -1.5) {};
		\node [style=Minor State, line width=0.5mm] (6) at (1, -1.5) {Ready};
		\node [style=none] (9) at (1.6, 1.5) {$\Delta t_\mathrm{takeoff}$};
		\node [style=none] (10) at (-1.6, 1.5) {$\Delta t_\mathrm{en-route}$};
		\node [style=none] (11) at (-2.5, -1.5) {$\Delta t_\mathrm{landing}$};
		\node [style=none] (12) at (-0.2, -2.1) {$\Delta t_\mathrm{turnaround}$};
		\node [style=none, align=left] (13) at (3.25, -1.75) {$\Delta t_\mathrm{idling}$ until operation\\request received};
		\node [style=none, align=left] (14) at (6.7, 0) {$\Delta t_\mathrm{ground-holding}$ until\\departure fix available};
		\node [style=none, align=left] (15) at (-5.9, 0) {$\Delta t_\mathrm{air-holding}$ until\\landing space\\and approach fix\\available};
		
		\node [style=rectangle, draw=black, line width=0.5mm, align=left] (15) at (7, -1.75) {Operation Center};
	\end{pgfonlayer}
	\begin{pgfonlayer}{edgelayer}
		\draw [->, >=stealth, black, thick, bend right=15, line width=0.5mm] (3) to (4);
		\draw [->, >=stealth, black, thick, in=-165, out=165, loop, line width=0.5mm] (2) to ();
		\draw [->, >=stealth, black, thick, in=15, out=-15, loop, line width=0.5mm] (3) to ();
		\draw [->, >=stealth, black, thick, bend right=15, line width=0.5mm] (4) to (2);
		\draw [->, >=stealth, black, thick, bend right=15, line width=0.5mm] (2) to (5);
		\draw [->, >=stealth, black, thick, bend right=15, line width=0.5mm] (5) to (6);
		\draw [->, >=stealth, black, thick, bend right=15, line width=0.5mm] (6) to (3);
		\draw [->, >=stealth, black, thick, dashed, line width=0.5mm] (15) to (13);
	\end{pgfonlayer}
\end{tikzpicture}
    \caption{Markov process  for UAM operation simulation.}
    \label{fig:simulation}
\end{figure}
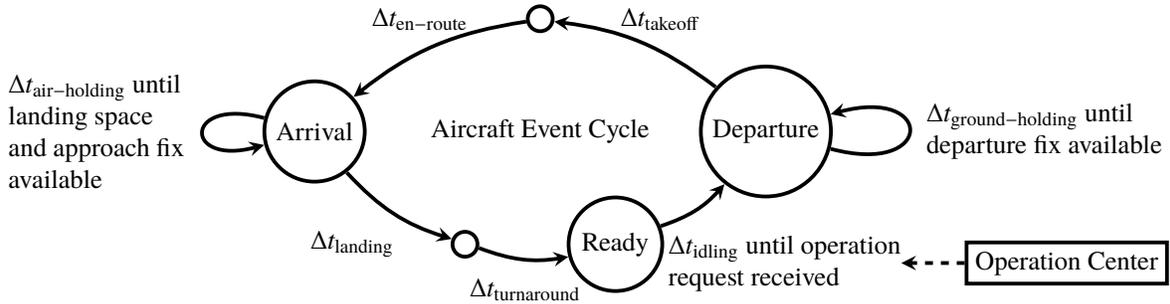

\cref{fig:simulation} shows the simulation event cycle for a single operation as a Markov chain.
In this formulation, an operation can be represented as a Markov chain (~\cref{fig:vertiport_operation}). 
 Starting from the ``Ready'' state, an aircraft idles for $\Delta t_\text{idling}$ until receiving an operation request. 
An operation request is generated by sampling a random distribution, and passed into the model through an operation center. 
Then the aircraft enters the ``Departure'' state, holding for $\Delta t_\text{ground-holding}$ until a departure slot is available. Then it takes-off with $\Delta t_\text{takeoff}$, and spends $\Delta t_\text{en-route}$ to fly towards the destination vertiport. Upon arrival at destination, the aircraft is held airborne for $\Delta t_\text{air-holding}$ until a landing space and an approach fix are available. Then it lands with $\Delta t_\text{landing}$. After spending $\Delta t_\text{turnaround}$ for turnaround procedures, the aircraft will be in the ``Ready'' state again. 

We make a number of simplifying assumptions to ensure that the event-driven simulation is tractable and can scale. 
In the context of our framework, these assumptions can be considered frozen design variables. 
We leave the analysis of their impact on the performance of the UAM ecosystem as future work. 
The assumptions are as follows:

\begin{enumerate}
\item Concept of operations and configurations of vertiports:
\begin{itemize}
    \item a single operator is using the airspace and vertiports in the region
    \item operations are in the nominal range, and disturbances such as weather and aircraft system failures are not considered
    \item dynamic interactions with manned traffic are not explicitly considered, with UAM operations assumed to use dedicated corridors that do not interfere with manned traffic
    \item all the vertiports and their properties are the same in a UAM network
    \item a vertiport has the ability to accept a single arrival or departure if available capacity exists
    \item preparation (turnaround of aircraft) for flight happens in a location in the vertiport separate from the landing and take-off zone
\end{itemize}

\item Airspace structure:
\begin{itemize}
    \item flight corridors are pre-defined in the UAM network to avoid collisions~\cite{thipphavong2018urban}
    \item no hard limit is placed on how many aircraft can be placed in an air-holding pattern at the same time
    \item the airspace of a vertiport has one approach fix and one departure fix
    \item appropriate separation requirements are assumed to be met when an aircraft is using the defined approach and departure fixes
\end{itemize}

\item Rules of sequencing and spacing:
\begin{itemize}
    \item ``first come, first served'' for demand dispatching, landing and takeoff sequencing
    \item separation is enforced pre-flight through time based deconfliction 
\end{itemize}
\end{enumerate}

\subsection{Time Utilization in the UAM Ecosystem}\label{sec:time_util}
The primary measures used to quantify the operational performance of the UAM ecosystem considered in this work are based on time utilization of the aircraft. 
The level of passenger satisfaction, the energy consumption requirements of the fleet, the monetary cost of the operation, and the operation capacity can all be expressed in terms of a time valued measure. 
Specifically, we consider the average time a vehicle spends in a given state or set of states as part of a nominal operation profile (as shown in~\cref{fig:simulation}).
The times of interest for this work are outlined in~\cref{tab:metrics} along with their definitions. 
We assume the time metrics are random variables, with expected values computed empirically using the statistical averages for those times in simulation.

\begin{table}[h]
    \centering
    \small
    \caption{Segments for time utilization analysis of a UAM ecosystem.}
    \label{tab:metrics}
    \begin{tabular}{lll}
        \toprule
        Metric  &Starting time &Ending time \\ \midrule
        $\Delta t_\text{demand delay}$ &Operation request submitted &Aircraft carrying out the request takes-off\\
        $\Delta t_\text{idling}$ &Aircraft is prepared for flight &Request is matched with a readied aircraft  \\
        $\Delta t_\text{ground-holding}$ &Aircraft is mission ready &Aircraft takes-off\\
        $\Delta t_\text{en-route}$ &Aircraft takes off &Aircraft lands at destination vertiport\\
        $\Delta t_\text{air-holding}$ &Aircraft enters the destination airspace &Aircraft starts landing\\
        $\Delta t_\text{re-balancing}$ &Re-balancing aircraft takes-off &Re-balancing aircraft lands\\
        \bottomrule
    \end{tabular}
\end{table}



\subsection{Stochastic Demand Models}
\begin{figure}[h!]
    \centering
    \input{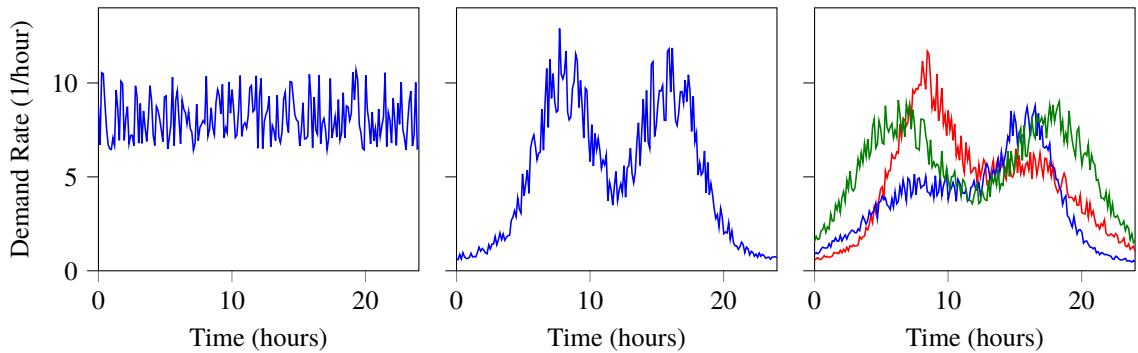}
    \caption{From left to right 1) temporally uniform demand model, 2) temporally modulated demand model using Gaussian mixture, and 3) geographically imbalanced demand models using Gaussian mixture for a three-vertiport UAM network, three different colors indicate the different demand profiles associated with each of vertiport.}
    \label{fig:demand_models}
\end{figure}

Operation requests are the driving force of the UAM ecosystem.
One way of modeling request generation is modeling it as a Poisson process. 
We can sample the number of requests generated in a time period from a Poisson distribution, which we refer to as the temporal demand model. 
A trivial model is a temporally uniform demand model where demand rate is time independent~\cite{goyal2018urban}. 
A uniform model is not realistic, but useful for examining the steady-state performance of a UAM ecosystem. 
Due to the time-varying nature of UAM demand, we are interested in modeling time dependent demand rate as well. 
A more realistic model includes a time modulated demand rate, where demand rate changes with respect to time. 
Using a Gaussian mixture model~\cite{kohlman2018system}, we are able to emulate peaks and lows in demand for UAM operations. 
We then generate operation requests using a time-varying Poisson process parameterized by demand rates sampled from the Gaussian mixture model. 
By shifting the center and height of Gaussian mixture peaks, we can create geographically imbalanced demand rates for different vertiports to emulate load imbalance among routes which is a common phenomenon seen in ground transportation. From left to right in \cref{fig:demand_models} shows examples of a temporally uniform demand model, temporally modulated demand model and imbalanced demand model for vertiports.

\section{System Performance Analysis}
\label{sec:sys-pef-analysis}

\begin{figure}[h!]
    \centering
    \input{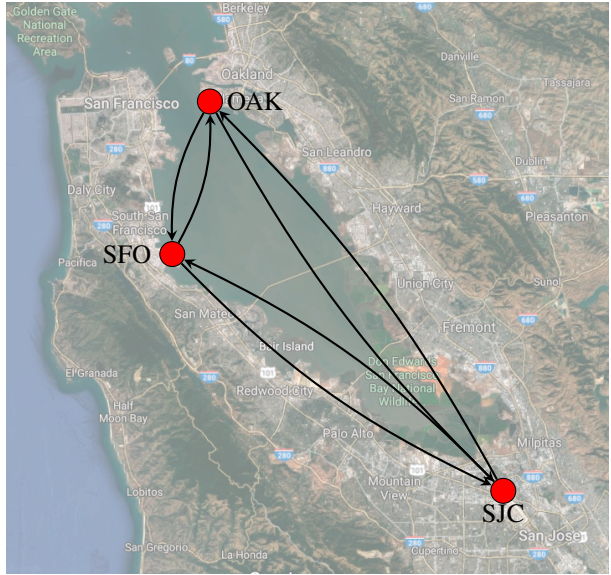}
    \caption{A conceptual UAM network in the San Francisco Bay Area where three vertiports are located near three major airports. All vertiports are connected.}
    \label{fig:bayairspace}
\end{figure}

In this work, we consider a conceptual UAM ecosystem in the San Francisco Bay Area. 
For simplicity, we assume there are three vertiports located near three major airports in the area (SFO, SJC and OAK). 
All the vertiports are directly connected by UAM routes, and are within the operational range of the UAM vehicles serving them on a single battery charge. 
We leave more complex vertiport configuration optimization, and route structure analyses for future work.
\cref{fig:bayairspace} shows the vertiport locations and connected routes in the UAM network. 
We configure the mean turnaround time to 10 min, with a standard deviation of 5 minutes based on average times a vehicle may spend between landing and being operation ready~\cite{vascik2019development}.
We set the average aircraft cruise speed to 140 km/h, with a nominal operational profile following the Markov process outlined in~\cref{fig:simulation}.
In this analysis, we fix the fleet size, but vary the vertiport constraints and the fleet management policy (see~\cref{tab:sys-peformance-vaiables}). 
We examine the delays, time utilization, and throughput metrics for time-independent and time-modulated demand models. 
Additionally, we analyze the results in the context of the UAM network capacity which is a function of fleet size and average operation time computed in~\cref{eqn:max_demand_est}, it is found to be $\hat{d}_\mathrm{max} = 67.5$ operations per hour for the configuration analyzed in this section. 
We consider how the performance metrics for the ecosystem vary as a function of demand, vertiport capacity, and traffic management policies.  
We evaluate the following three traffic management policies: an on-demand policy without fleet re-balancing, an on-demand policy with fleet re-balancing (see~\cref{sec:rebalancing}, and a policy that follows a fixed schedule (see~\cref{sec:scheduling}). 
We also examine the impact of normalized vertiport capacity on the performance metrics.  
The goal of the analysis in this section is quantify how choice of vertiport capacity and fleet management policy impact system performance. 

\begin{table}[]
\caption{Design variables used in the system performance analysis.}
\label{tab:sys-peformance-vaiables}
\centering
\small
    \begin{tabular}{lccc}
        \toprule
          & Vertiport Capacity & Fleet Size & Fleet Management Policy  \\ \midrule
        Variable Value & Varied & 36 & Varied \\
        \bottomrule
    \end{tabular}
\end{table}

\subsection{Temporally Uniform Demand Model}\label{sec:anaylsis_uniform}

\begin{figure}[h!]
    \centering
    \input{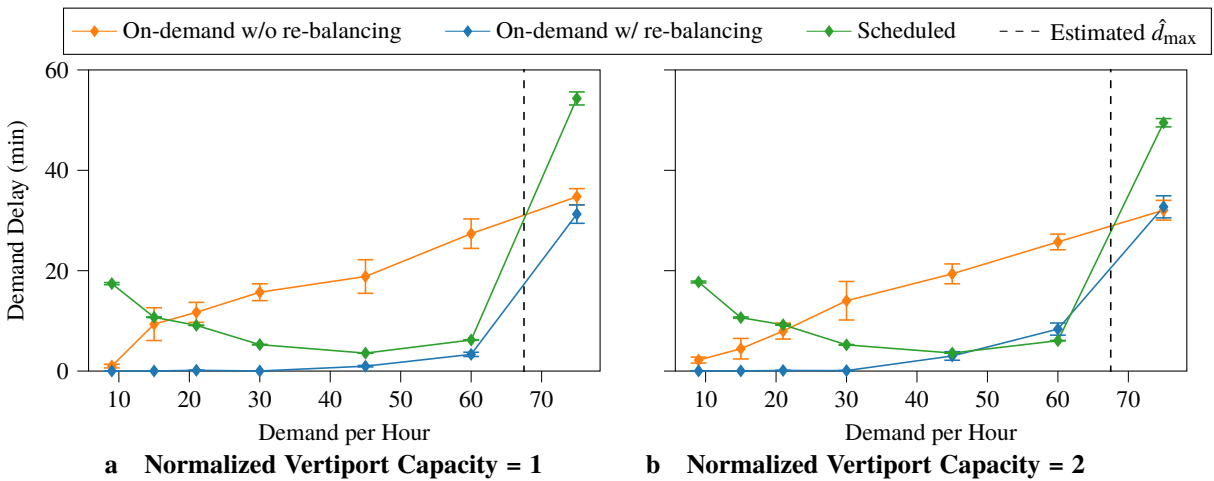}
    \caption{Demand delay analysis of UAM ecosystems with normalized vertiport capacity set to 1 and 2 under a uniform demand model.}
    \label{fig:demand_delay_uniform}
\end{figure}

First, we analyze operational performance under a time-independent uniform demand model where the average demand rate is a constant with respect to time.
\cref{fig:demand_delay_uniform} shows the demand delay, or the time delay between the desired and the actual start times of an operation as the average demand rate in the ecosystem increases. 
We see little impact of increasing vertiport capacity on the demand delay metric in the emergent system.  
However, the choice of fleet management policy has significant effects on the delay metric.
We observe that the on-demand + re-balancing policy leads to shorter demand delays at all demand rates evaluated. 
This is expected as the policy accounts for the stochastic nature of passenger demand in the system, and attempts to uniformly distribute the UAM fleet reducing delay times.  
A lack of a re-balancing policy leads to larger delays, as an imbalanced fleet can both create congestion at vertiports where vehicles are idle and lack of readied vehicles for operations when requests come into the system.  
The relatively large delays at lower demand rates for scheduled operations are surprising, and lead to a U shaped curve for the scheduled fleet management policy.  
This can be explained by the very low frequency of scheduled flights that occur at lower demand rates. 
There is a natural analogy in ground transportation in for this phenomenon. 
A bus schedule in rural areas leads to relatively infrequent bus runs, and showing up at a random time of day to ride it can lead to long wait times; however the bus schedule in busy urban area can be lead to more frequent bus runs, leading to a shorter wait time for passengers.

\begin{figure}[h!]
    \centering
    \input{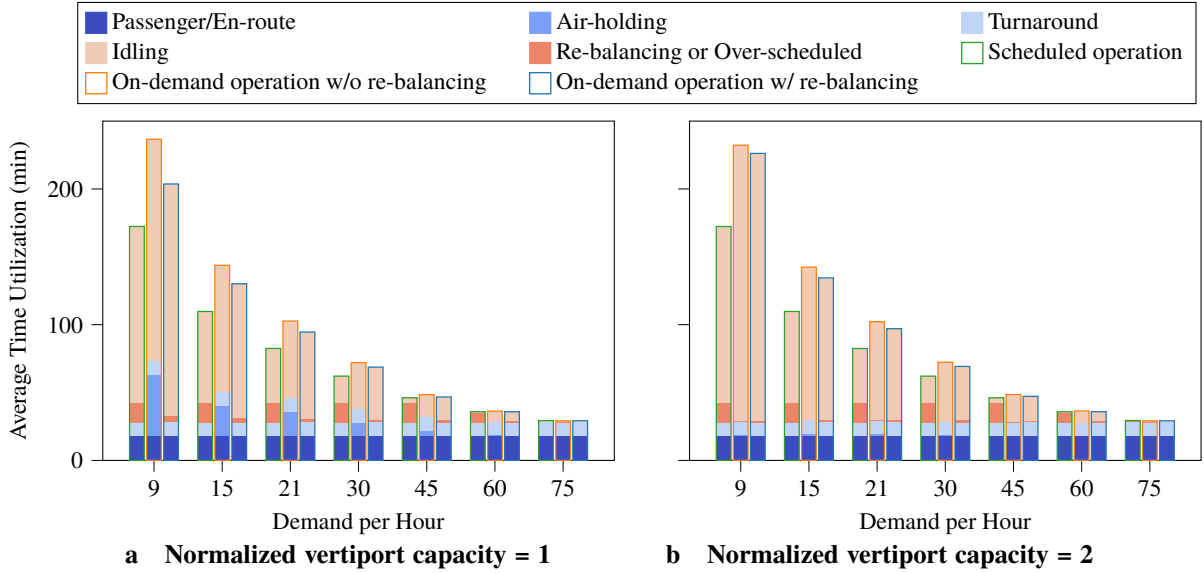}
    \caption{Time utilization analysis of UAM ecosystems with normalized vertiport capacity of 1 (left) and 2 (right) under a uniform demand model.}
    \label{fig:time_util_uniform}
\end{figure}

\cref{fig:time_util_uniform} shows a bar plot of time utilization for the three fleet management policies evaluated in this work as a function of demand. 
We use time metrics introduced in~\cref{sec:time_util} to compare how much time a vehicle spends in various states between operations. 
Specifically, we analyze the time a vehicle spends en-route, air-holding, idling, in a maintenance or turnaround state, and in a re-balancing or over-scheduled state. 
Time spent re-balancing or over-scheduled implies that a vehicle is flying empty to either re-balance the fleet or to perform its scheduled flight without a passenger.
Larger normalized vertiport capacity significantly reduces and in certain cases eliminates the air-holding time of the on-demand policy with no re-balancing.
This occurs because additional vertiport capacity allows more vehicles to idle at a vertiport.
Without re-balancing and with limited vertiport capacity, en-route operations may enter an air-holding pattern when there is no room to accommodate them at the destination. 
Without re-balancing, the only mechanism that creates room for an en-route operation in a full vertiport is an outgoing operation.
We note that increased vertiport capacity has negligible impacts on the other two fleet management policies. 
The results also show that fleet re-balancing is an effective way of preventing air-holding by freeing up used capacity at congested vertiports.
As the demand rate increases, air-holding time and re-balancing time both decrease. 
This occurs because a high demand rate drives traffic flow, and in a uniform demand setting can naturally re-balance the fleet.
\textit{Fleet re-balancing creates additional traffic flow that can geographically re-distribute a UAM fleet in a more uniform way}. 
We note that when the demand rate exceeds the estimated capacity from~\cref{eqn:max_demand_est}, idling trends towards zero. 
At this stage the fleet is being fully utilized by either carrying passenger or being prepped for carrying passengers in the turnaround state. 
We note that while fleet utilization is maximized, on an operation by operation basis, passengers experience large delays once the capacity of the UAM ecosystem is exceeded.

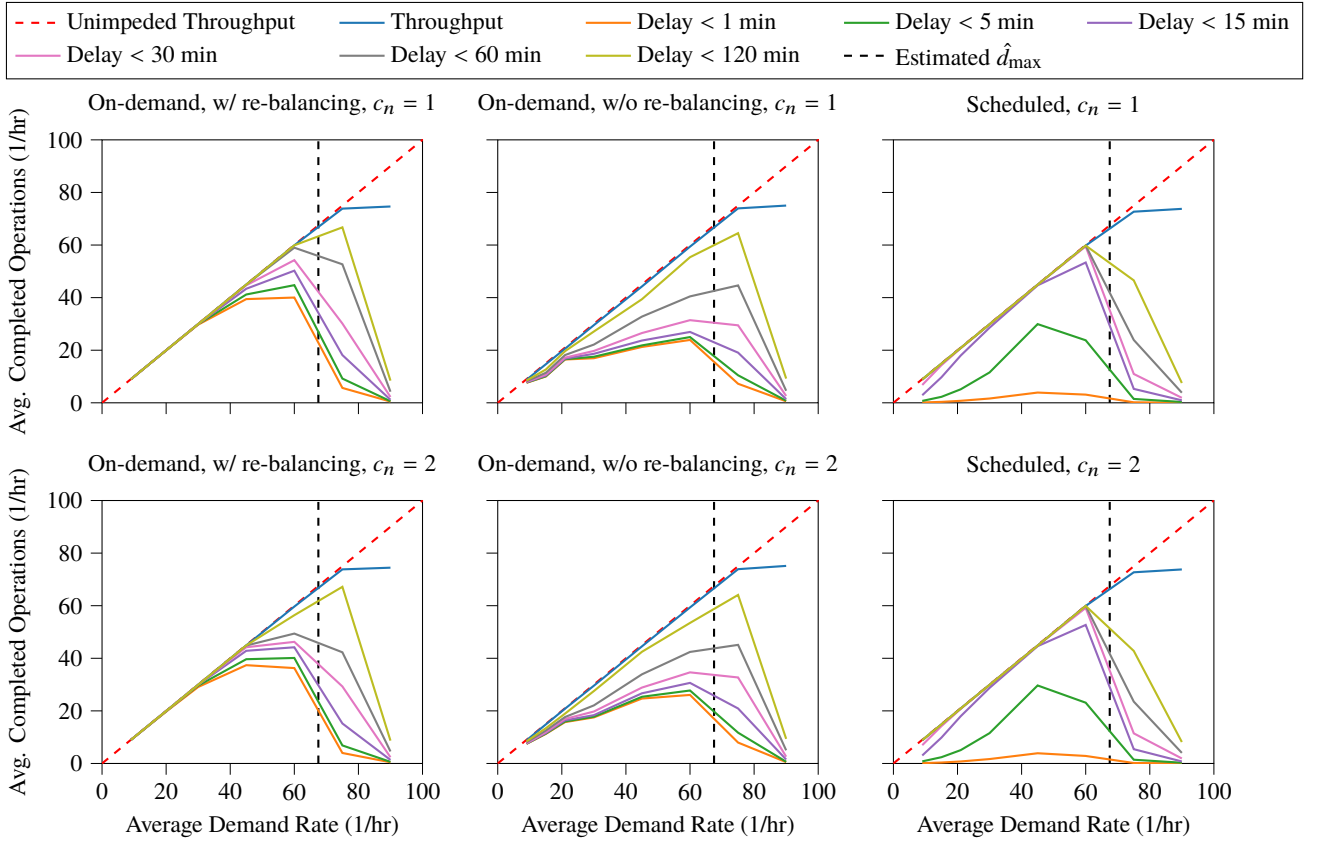
\begin{figure}[h!]
    \centering
    \begin{tikzpicture}
\small
\definecolor{color0}{rgb}{0.12156862745098,0.466666666666667,0.705882352941177}
\definecolor{color1}{rgb}{1,0.498039215686275,0.0549019607843137}
\definecolor{color2}{rgb}{0.172549019607843,0.627450980392157,0.172549019607843}
\definecolor{color3}{rgb}{0.83921568627451,0.152941176470588,0.156862745098039}
\definecolor{color4}{rgb}{0.580392156862745,0.403921568627451,0.741176470588235}
\definecolor{color5}{rgb}{0.549019607843137,0.337254901960784,0.294117647058824}
\definecolor{color6}{rgb}{0.890196078431372,0.466666666666667,0.76078431372549}
\definecolor{color7}{rgb}{0.737254901960784,0.741176470588235,0.133333333333333}

\begin{groupplot}[
group style={columns=3,
        rows=2,
        group name=plots,
        x descriptions at=edge bottom,
        y descriptions at=edge left,
        vertical sep=1.3cm}, 
tick align=outside,
tick pos=left,
x grid style={white!69.01960784313725!black},
xlabel={Average Demand Rate (1/hr)},
xmin=0, xmax=100,
xtick style={color=black},
y grid style={white!69.01960784313725!black},
ylabel={Avg. Completed Operations (1/hr)},
ymin=0, ymax=100,
ytick style={color=black},
width={2.3in}, height={2in},
legend columns=5,
legend cell align={left},
legend style={
    at={(-0.3,1.23)}, 
    /tikz/every even column/.append style={column sep=0.5cm},
    anchor={south west}, 
    nodes={scale=1.0, transform shape}
},
]

\nextgroupplot[title={On-demand, w/ re-balancing, $c_{n}=1$}]
\addlegendimage{mark=--, thick, dashed, color=red}\addlegendentry{Unimpeded Throughput}
\addlegendimage{thick, color=color0}\addlegendentry{Throughput}
\addlegendimage{thick, color=color1}\addlegendentry{Delay $<$ 1 min}
\addlegendimage{thick, color=color2}\addlegendentry{Delay $<$ 5 min}
\addlegendimage{thick, color=color4}\addlegendentry{Delay $<$ 15 min}
\addlegendimage{thick, color=color6}\addlegendentry{Delay $<$ 30 min}
\addlegendimage{thick, color=white!49.80392156862745!black}\addlegendentry{Delay $<$ 60 min}
\addlegendimage{thick, color=color7}\addlegendentry{Delay $<$ 120 min}
\addlegendimage{thick, color=black, dashed}\addlegendentry{Estimated $\hat{d}_\text{max}$}

\addplot [thick, red, dashed]
table {%
0 0
200 200
};
\addplot [thick, black, dashed]
table {%
67.5 0
67.5 200
};
\addplot [thick, color0]
table {%
9 8.954
15 14.972
21 20.966
30 29.934
45 44.946
60 59.894
75 73.848
90 74.658
};
\addplot [thick, color1]
table {%
9 8.918
15 14.936
21 20.862
30 29.726
45 39.438
60 40.03
75 5.658
90 0.472
};
\addplot [thick, color2]
table {%
9 8.954
15 14.972
21 20.966
30 29.908
45 41.194
60 44.742
75 9.212
90 0.604
};
\addplot [thick, color4]
table {%
9 8.954
15 14.972
21 20.966
30 29.934
45 43.372
60 50.244
75 18.148
90 1.16
};
\addplot [thick, color6]
table {%
9 8.954
15 14.972
21 20.966
30 29.934
45 44.688
60 54.25
75 30.144
90 2.21
};
\addplot [thick, white!49.80392156862745!black]
table {%
9 8.954
15 14.972
21 20.966
30 29.934
45 44.946
60 58.99
75 52.672
90 4.218
};
\addplot [thick, color7]
table {%
9 8.954
15 14.972
21 20.966
30 29.934
45 44.946
60 59.894
75 66.724
90 8.43
};

\nextgroupplot[title={On-demand, w/o re-balancing, $c_{n}=1$}]
\addplot [thick, red, dashed]
table {%
0 0
200 200
};
\addplot [thick, black, dashed]
table {%
67.5 0
67.5 200
};
\addplot [thick, color0]
table {%
9 8.786
15 14.656
21 20.566
30 29.538
45 44.248
60 59.388
75 73.942
90 75.01
};
\addplot [thick, color1]
table {%
9 7.608
15 10.014
21 16.47
30 16.956
45 21.244
60 23.908
75 7.242
90 0.566
};
\addplot [thick, color2]
table {%
9 7.684
15 10.1
21 16.59
30 17.506
45 21.832
60 24.998
75 10.466
90 0.784
};
\addplot [thick, color4]
table {%
9 7.806
15 10.368
21 16.818
30 18.63
45 23.682
60 26.948
75 19.074
90 1.4
};
\addplot [thick, color6]
table {%
9 7.972
15 10.728
21 17.18
30 19.726
45 26.536
60 31.428
75 29.468
90 2.598
};
\addplot [thick, white!49.80392156862745!black]
table {%
9 8.214
15 11.41
21 18.156
30 22.14
45 32.784
60 40.468
75 44.638
90 4.582
};
\addplot [thick, color7]
table {%
9 8.518
15 12.644
21 19.636
30 27.112
45 39.392
60 55.372
75 64.508
90 9.216
};

\nextgroupplot[title={Scheduled, $c_{n}=1$}]
\addplot [thick, red, dashed]
table {%
0 0
200 200
};
\addplot [thick, black, dashed]
table {%
67.5 0
67.5 200
};
\addplot [thick, color0]
table {%
9 8.936
15 14.906
21 20.872
30 29.856
45 44.904
60 59.794
75 72.682
90 73.76
};
\addplot [thick, color1]
table {%
9 0.138
15 0.35
21 0.762
30 1.63
45 3.898
60 3.114
75 0.188
90 0.058
};
\addplot [thick, color2]
table {%
9 0.716
15 2.276
21 5.138
30 11.554
45 29.932
60 23.778
75 1.47
90 0.312
};
\addplot [thick, color4]
table {%
9 2.888
15 9.73
21 17.974
30 28.57
45 44.672
60 53.354
75 5.292
90 0.938
};
\addplot [thick, color6]
table {%
9 6.782
15 14.328
21 20.74
30 29.838
45 44.904
60 59.47
75 10.968
90 1.896
};
\addplot [thick, white!49.80392156862745!black]
table {%
9 8.738
15 14.892
21 20.872
30 29.856
45 44.904
60 59.794
75 23.898
90 3.844
};
\addplot [thick, color7]
table {%
9 8.936
15 14.906
21 20.872
30 29.856
45 44.904
60 59.794
75 46.562
90 7.526
};

\nextgroupplot[title={On-demand, w/ re-balancing, $c_{n}=2$}]
\addplot [thick, red, dashed]
table {%
0 0
200 200
};
\addplot [thick, black, dashed]
table {%
67.5 0
67.5 200
};
\addplot [thick, color0]
table {%
9 8.96
15 14.96
21 20.92
30 29.932
45 44.9
60 59.664
75 73.814
90 74.454
};
\addplot [thick, color1]
table {%
9 8.944
15 14.816
21 20.66
30 29.086
45 37.38
60 36.276
75 3.986
90 0.452
};
\addplot [thick, color2]
table {%
9 8.95
15 14.896
21 20.848
30 29.6
45 39.652
60 40.122
75 6.856
90 0.614
};
\addplot [thick, color4]
table {%
9 8.96
15 14.954
21 20.92
30 29.916
45 42.848
60 44.184
75 15.21
90 1.348
};
\addplot [thick, color6]
table {%
9 8.96
15 14.96
21 20.92
30 29.932
45 44.202
60 46.248
75 29.284
90 2.356
};
\addplot [thick, white!49.80392156862745!black]
table {%
9 8.96
15 14.96
21 20.92
30 29.932
45 44.846
60 49.398
75 42.28
90 4.56
};
\addplot [thick, color7]
table {%
9 8.96
15 14.96
21 20.92
30 29.932
45 44.9
60 56.384
75 67.182
90 8.682
};

\nextgroupplot[title={On-demand, w/o re-balancing, $c_{n}=2$}]
\addplot [thick, red, dashed]
table {%
0 0
200 200
};
\addplot [thick, black, dashed]
table {%
67.5 0
67.5 200
};
\addplot [thick, color0]
table {%
9 8.836
15 14.836
21 20.832
30 29.574
45 44.536
60 59.332
75 73.89
90 75.134
};
\addplot [thick, color1]
table {%
9 7.456
15 11.228
21 15.734
30 17.52
45 24.688
60 26.06
75 7.948
90 0.504
};
\addplot [thick, color2]
table {%
9 7.502
15 11.354
21 15.958
30 17.768
45 25.356
60 27.746
75 11.708
90 0.65
};
\addplot [thick, color4]
table {%
9 7.558
15 11.568
21 16.408
30 18.37
45 26.682
60 30.642
75 20.91
90 1.504
};
\addplot [thick, color6]
table {%
9 7.706
15 11.904
21 16.922
30 19.754
45 28.862
60 34.648
75 32.706
90 2.578
};
\addplot [thick, white!49.80392156862745!black]
table {%
9 8.07
15 12.53
21 17.652
30 22.08
45 33.898
60 42.414
75 45.102
90 5.002
};
\addplot [thick, color7]
table {%
9 8.436
15 13.39
21 18.954
30 27.534
45 42.466
60 53.444
75 64.092
90 9.36
};

\nextgroupplot[title={Scheduled, $c_{n}=2$}]
\addplot [thick, red, dashed]
table {%
0 0
200 200
};
\addplot [thick, black, dashed]
table {%
67.5 0
67.5 200
};
\addplot [thick, color0]
table {%
9 8.9
15 14.858
21 20.914
30 29.892
45 44.812
60 59.902
75 72.7
90 73.788
};
\addplot [thick, color1]
table {%
9 0.126
15 0.354
21 0.76
30 1.676
45 3.886
60 2.858
75 0.176
90 0.05
};
\addplot [thick, color2]
table {%
9 0.794
15 2.388
21 5.076
30 11.532
45 29.664
60 23.112
75 1.42
90 0.282
};
\addplot [thick, color4]
table {%
9 3.034
15 9.848
21 17.97
30 28.736
45 44.608
60 52.688
75 5.424
90 0.758
};
\addplot [thick, color6]
table {%
9 6.954
15 14.214
21 20.746
30 29.878
45 44.812
60 59.146
75 11.45
90 1.896
};
\addplot [thick, white!49.80392156862745!black]
table {%
9 8.738
15 14.852
21 20.914
30 29.892
45 44.812
60 59.902
75 23.516
90 4.024
};
\addplot [thick, color7]
table {%
9 8.9
15 14.858
21 20.914
30 29.892
45 44.812
60 59.902
75 42.796
90 8.152
};
\end{groupplot}
\end{tikzpicture}
    \caption{Throughput analysis for the temporally uniform demand model with normalized vertiport capacity $c_{n}=1$ and $c_{n}=2$.}
    \label{fig:throughput_uniform}
\end{figure}

\cref{fig:throughput_uniform} shows the throughput analysis for the temporally uniform demand model. 
Throughput is defined as the average hourly rate of completed operations in the ecosystem. 
The line marking unimpeded throughput indicates a throughput rate that is equal to the demand rate.
The figure also shows per hour rates for operations completed within a specific delay window ranging from 1 minute to 2 hours.  
From~\cref{fig:throughput_uniform}, we observe that normalized vertiport capacity has little impact on net average delay and the overall throughput of the system. 
Additionally, the average throughput begins to flatten beyond $\sim 75$ operation requests per hour for all of the fleet management policies, which coincides with the estimated capacity bound for the UAM network, which was found to be 67.5 operations per hour.
However, the delay performance differs drastically between fleet management policies. 
Specifically, we observe increasing system-wide delays as the management policy changes from re-balancing, to non re-balancing, to scheduled.

\subsection{Temporally Modulated Gaussian Mixture Demand Model}
\begin{figure}[h!]
    \centering
    \input{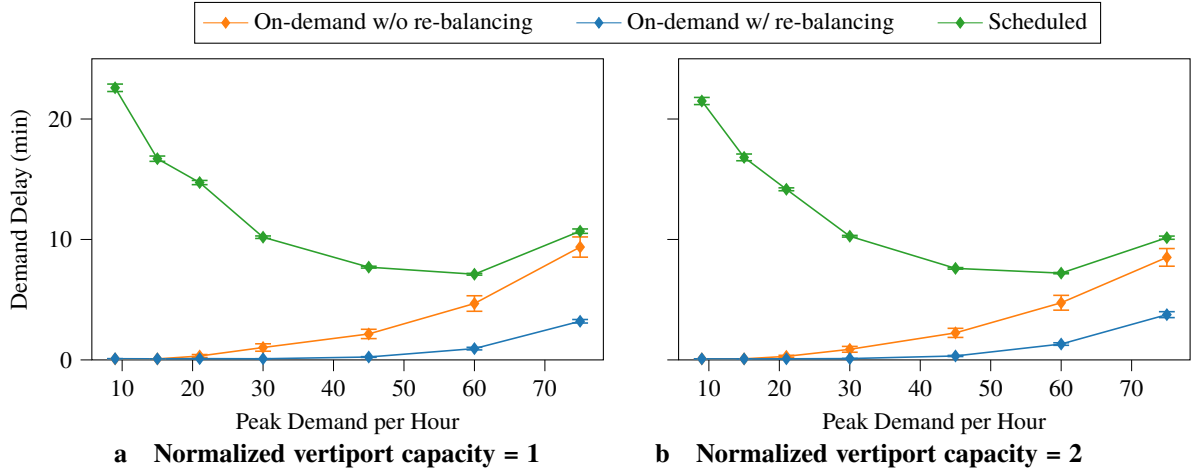}
    \caption{Demand delay analysis of UAM ecosystems with normalized vertiport capacity of 1 and 2 under Gaussian mixture demand model.}
    \label{fig:demand_delay_gmm}
\end{figure}

We perform similar analysis on a temporally modulated Gaussian mixture demand model with equally weighted components where the time varying demand is proportional to $d(t) \propto \text{Normal}(t; 8, 2) + \text{Normal}(t; 12, 8) + \text{Normal}(t; 16, 2)$ where $t$ is given in hours~\cite{kohlman2018system}. 
This demand model simulates peak hours at 8 AM and 4 PM, moderate demand rate at noon and close to zero demand rate at midnight. 
The middle figure in \cref{fig:demand_models} illustrates this Gaussian mixture model with 25\% noise. 
We control the total demand rate by adjusting the peak demand rate of the model, and it is with respect to the peak demand rate that the results are presented. 
\cref{fig:demand_delay_gmm} shows the demand delay analysis of the conceptual UAM ecosystem with normalized vertiport capacity of $c_{n}=1$ and $c_{n}=2$. 
A higher vertiport capacity is not shown to have major impacts on the delays. 
Additionally, we observe an improvement in the performance of the fleet management policy without re-balancing and a worsened delay performance for the fixed schedule fleet management policy. 

\begin{figure}[h!]
    \centering
    \input{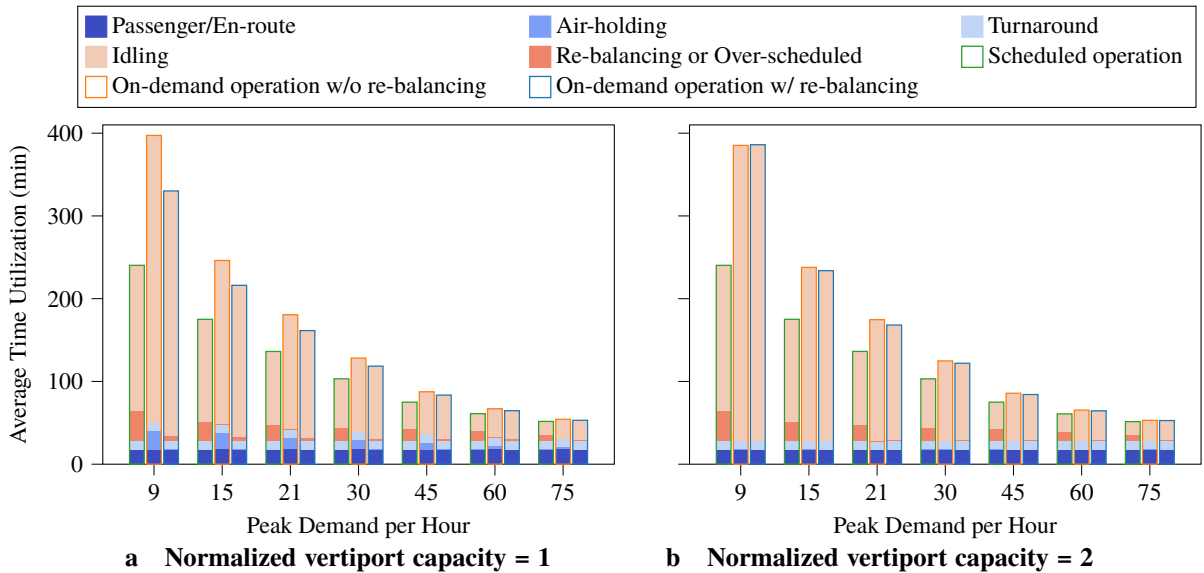}
    \caption{Time utilization analysis of UAM ecosystems with normalized vertiport capacity of 1 and 2 under a Gaussian mixture demand model}
    \label{fig:time_util_gmm}
\end{figure}

\cref{fig:time_util_gmm} shows the time utilization as a function of peak demand rate. 
Similar to the results under a uniform demand model, a larger vertiport capacity significantly reduces air-holding time for on-demand operation without re-balancing, and we observe the policy with re-balancing avoiding air-delays while operating fewer empty flights than a fixed schedule policy. 
While the fixed schedule policy reduces idling time, it leads to the largest faction of operations flying empty pre-scheduled flights.
In fact, a time varying demand model leads to a higher ratio of empty flights compared to the ratio of empty flights for a uniform demand model. 
Naive fixed scheduling approaches will be difficult to apply efficiently to dynamic demand profiles, and we see significant gains in operating an on-demand policy in this work.
However, it is possible that demand may adjust to scheduled operations which could improve performance.
While it is possible that system performance under a scheduling policy can be improved, further analysis of scheduling optimization for UAM operations is left as future work.

\begin{figure}[h!]
    \centering
    \input{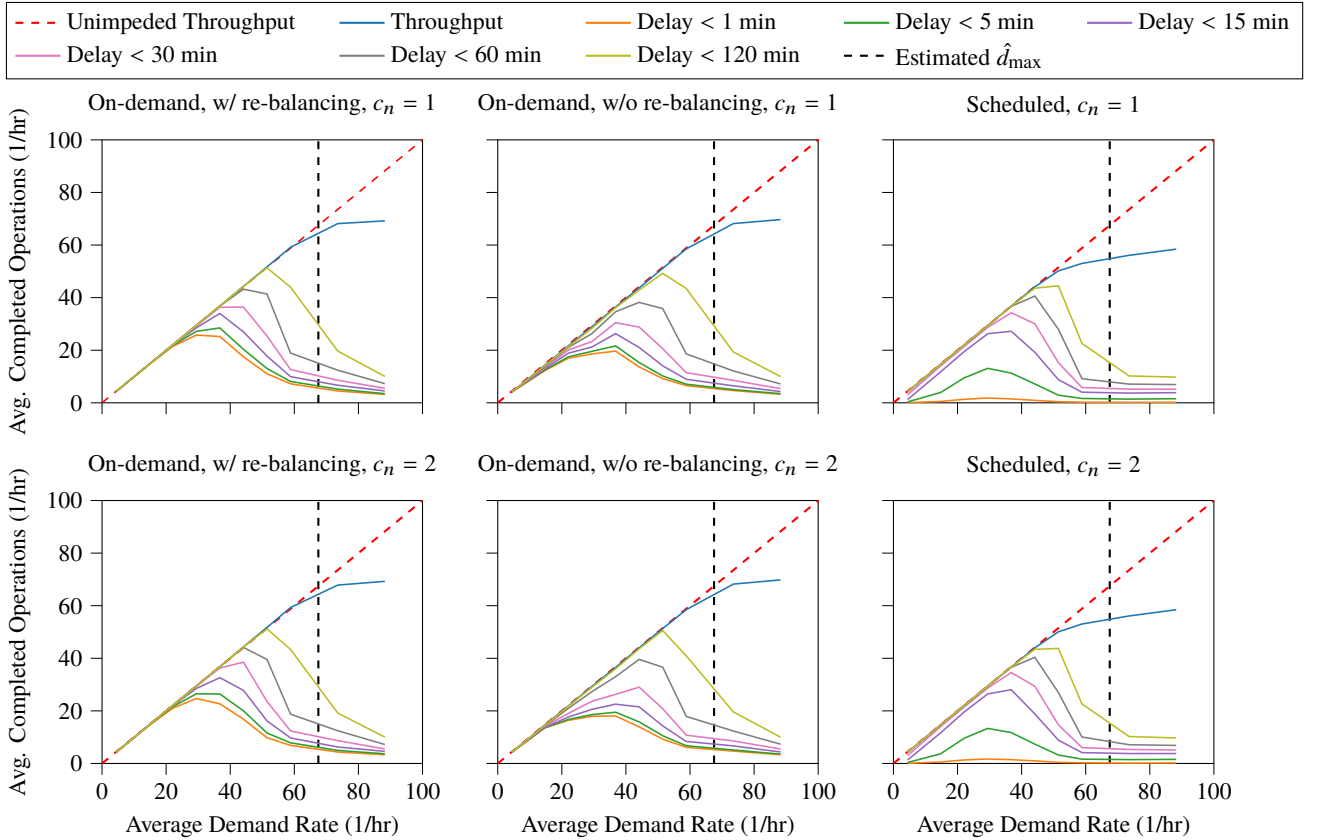}
    \caption{Throughput analysis for the temporally modulated Gaussian mixture demand model with normalized vertiport capacity $c_{n}=1$ and $c_{n}=2$.}
    \label{fig:throughput_gmm}
\end{figure}

\cref{fig:throughput_gmm} shows the throughput analysis for the temporally modulated Gaussian mixture demand model.
Similar to the results for uniform demand model, vertiport capacity does not play a significant role on average net delays. 
The on-demand policy with re-balancing shows the best throughput profile overall with a large fraction of operations seeing very low delays ($<$ 1 min).
The on-demand policy without re-balancing shows a high proportion of operations experience long delay ($>$ 30 min) as average demand rate increases over 30 per hour. 
The majority of operations under a fixed schedule policy tend to fall within delays 5 to 15 minutes.
The throughput values for all policies plateau and degrade at lower demand rates than those in temporally uniform demand analysis. 
This is because the temporally nonuniform demand rate can be low at non-peak hours and large at peak hours. 
The peak hour demand rate is higher than the average demand rate (around twice of average in this particular Gaussian mixture demand model).
The demand peaks can cause cascading effects that makes the UAM ecosystem reach its capacity at a lower average demand rate than when it is operated with a temporally uniform demand model. 
The estimate of capacity given by~\cref{eqn:max_demand_est} is therefore an over-estimate for a more realistic time varying demand model. 

\section{Ecosystem Design Optimization}

In this section, we perform UAM design variable optimization with respect to a cost function that is based on system wide performance metrics.
We use a similar simulation configuration as in~\cref{sec:sys-pef-analysis}, with a three vertiport network, operations cruising at 140 km/h, and mean turnaround time at vertiports of 10 minutes. 
However, we fix the fleet management policy to be the best performing policy we found in the previous section, on demand with re-balancing. 
We vary both the fleet size and the vertiport capacity. 
The goal of this analysis is to determine the sensitivities of system performance to strategic design variables like vertiport capacity and fleet size, when a performant fleet management policy is used (see~\cref{tab:sys-optimization-vaiables}).

\begin{table}[]
\caption{Design variables used in the ecosystem design optimization analysis.}
\label{tab:sys-optimization-vaiables}
\centering
\small
    \begin{tabular}{lccc}
        \toprule
          & Vertiport Capacity & Fleet Size & Fleet Management Policy  \\ \midrule
        Variable Value & Varied & Varied & On-Demand + Re-balancing \\
        \bottomrule
    \end{tabular}
\end{table}

\subsection{Cost Function Design}
A cost aware decision process has been used in a number of application in aviation, including in green airline fleet planning~\cite{tsai2012mixed}. 
We extend the formulation to consider variables and metrics of interest for UAM. 
We denote the cost $J_\mathrm{marginal}$, which quantifies the overall performance of the UAM ecosystem, and use it to guide the choice of design variables relevant to UAM.
We consider the cost $J_\mathrm{marginal}$ to be  ``marginal'' because it is the result of subtracting the nominal operation cost from the total operation cost, where the nominal operation cost only includes the basic cost for fulfilling the demand that exists within the ecosystem.
The cost from any unfulfilled demand is reflected in the cost associated with delays.
Additional cost induced by re-balancing or over-scheduled flights is also included in the cost function. 
Idling cost is added to represent the cost for fleet maintenance.
The cost function $J_\mathrm{marginal}$ is defined as follows

\begin{equation}\label{eq:cost}
\begin{aligned}
J_\mathrm{marginal} = ( &c_\text{idling} T_\text{tot, idling} + c_\text{ground-holding} T_\text{tot, ground-holding} + c_\text{air- holding} T_\text{tot, air-holding} \\
& + c_\text{cruising} T_\text{tot, additional cruising} + c_\text{takeoff} T_\text{tot, additional takeoff} + c_\text{landing} T_\text{tot, additional landing} \\
& + c_\text{demand delay} T_\text{tot, demand delay}) / T_\text{simulation},
\end{aligned}
\end{equation}
where the $c$'s represent various cost coefficients or weights whose values are either roughly proportional to energy consumption rates of different phases of operation~\cite{kohlman2018system}, or to a rough estimate of the cost of incurred delays as in $c_\text{demand delay}$. 
Subscript ``additional'' indicates that time is only accumulated for additional flights made in addition to customer requests (re-balancing or over-scheduled flights).
Time $T_\text{simulation}$ stands for the total simulation time duration. 
The weight assignments of cost coefficients are list in \cref{tab:cost_weights}.  

\begin{table}[h]
    \centering
    \caption{Weight of cost coefficients}
    \label{tab:cost_weights}
    \begin{tabular}{cc}
        \toprule
         Cost coefficients & Weight per time step \\ \midrule
         $c_\text{idling}$ &  1\\
         $c_\text{ground-holding}$ & 5\\
         $c_\text{air-holding}$ & 100\\
         $c_\text{cruising}$ & 50\\
         $c_\text{takeoff}$ & 150\\
         $c_\text{landing}$ & 150\\
         $c_\text{demand delay}$ & 100\\
         \bottomrule
    \end{tabular}
\end{table}

\subsection{Ecosystem Design Optimization}

\begin{figure}[h!]
    \centering
    \input{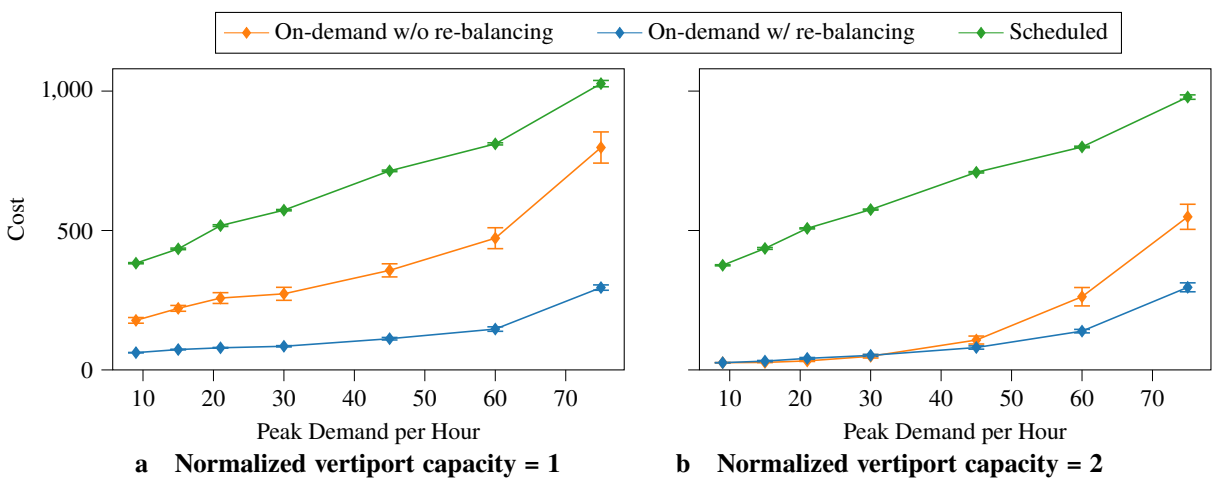}
    \caption{Marginal operation cost analysis of UAM ecosystem with normalized vertiport capacity of 1 and 2 unde a temporally modulated Gaussian mixture demand model.}
    \label{fig:cost_gmm}
\end{figure}

We first evaluate the cost as a function of demand, and show only the results for the time varying demand model for brevity. 
\cref{fig:cost_gmm} shows the marginal operation cost as a function of peak demand with normalized vertiport capacity of 1 and 2. 
Overall, on-demand operation with fleet re-balancing persistently performs better then the rest two operation rules. 
In \cref{fig:cost_gmm}b, on-demand operation without re-balancing performs as good as re-balanced on-demand operation under high normalized vertiport capacity up until peak demand rate reaching 45 per hour. 
\textit{We again find that increasing normalized vertiport capacity leads to improved performance when no fleet re-balancing exists, but is not significantly effective when fleets are managed with a re-balancing policy or are under a schedule.}
In the remainder of the section we consider on the on-demand with re-balancing fleet management policy as it is shown to be the most pefromant. 

We now examine, how the strategic UAM ecosystem design variables can be chosen in a way that minimizes the marginal operation cost $J_\text{marginal}$.
We fix the fleet management policy which is a tactical design variable, and vary both the vertiport capacity, and the fleet size in this analysis. 
Fixing the tactical design variable, allows a simplification the general optimization formulation~\cref{eqn:optim_general} to a reduced optimization problem \cref{eqn:optim}.
\begin{equation}\label{eqn:optim}
\begin{aligned}
& \underset{f, \mathcal{V}}{\text{minimize}}
& & J_\mathrm{marginal}(f, \mathcal{V},\Pi, \mathcal{N}, \mathcal{D}) \\
\end{aligned}
\end{equation}
Using simulation, we can approximately evaluate the whole design space using a discretized grid search. 
The vertiport configuration $\mathcal{V}$ we optimize over is the normalized vertiport capacity $c_{n}$. We jointly optimize the cost with respect to fleet size $f$. We search over $f\in\{15, 18, 21, \dots, 60\}$ and $c_{n} \in \{1.0, 0.1, \dots, 3.0\}$.

\begin{figure}[h!]
    \begin{subfigure}{.5\textwidth}
        \centering
        \includegraphics[trim={1cm 0 0 0}, clip, width=\textwidth]{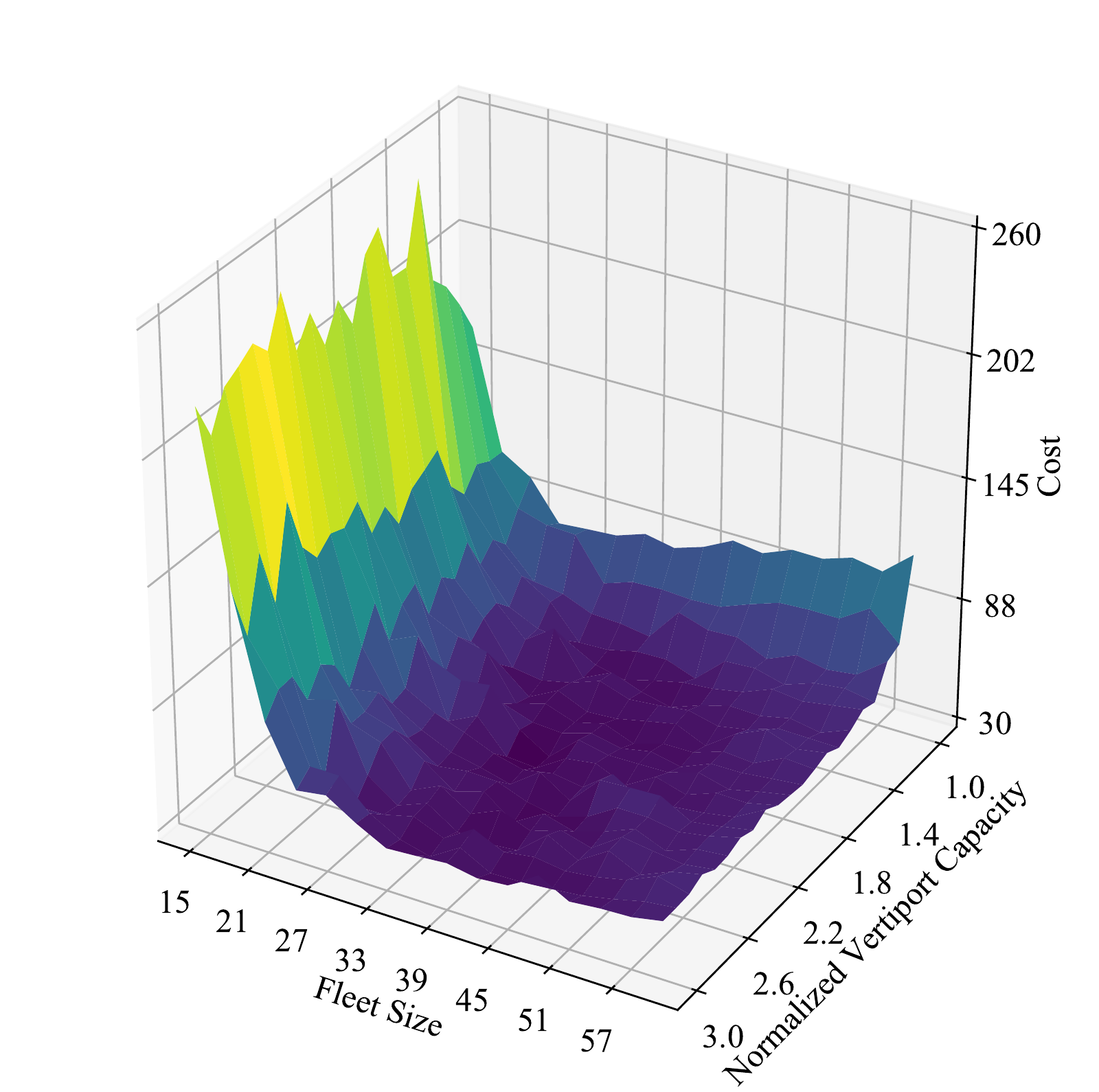}
        \caption{3D surface plot.}
        \label{fig:3d_surface}
    \end{subfigure}
    \hspace{0.1cm}
    \begin{subfigure}{.5\textwidth}
        \centering
        \small
\begin{tikzpicture}

\begin{axis}[
colorbar,
colorbar style={ylabel={Cost}},
colormap/viridis,
point meta max=250.2864,
point meta min=35.9086,
tick align=outside,
tick pos=left,
x grid style={white!69.01960784313725!black},
xlabel={Fleet Size},
xmin=-0.5, xmax=15.5,
xtick style={color=black},
xtick={0,2,4,6,8,10,12,14},
xticklabels={15,21,27,33,39,45,51,57},
y dir=reverse,
y grid style={white!69.01960784313725!black},
ylabel={Normalized Vertiport Capacity},
ymin=-0.5, ymax=20.5,
ytick style={color=black},
ytick={0,2,4,6,8,10,12,14,16,18,20},
yticklabels={3.0,2.8,2.6,2.4,2.2,2.0,1.8,1.6,1.4,1.2,1.0},
width=2.2in,
height=2.89in,
]
\addplot graphics [includegraphics cmd=\pgfimage,xmin=-0.5, xmax=15.5, ymin=20.5, ymax=-0.5] {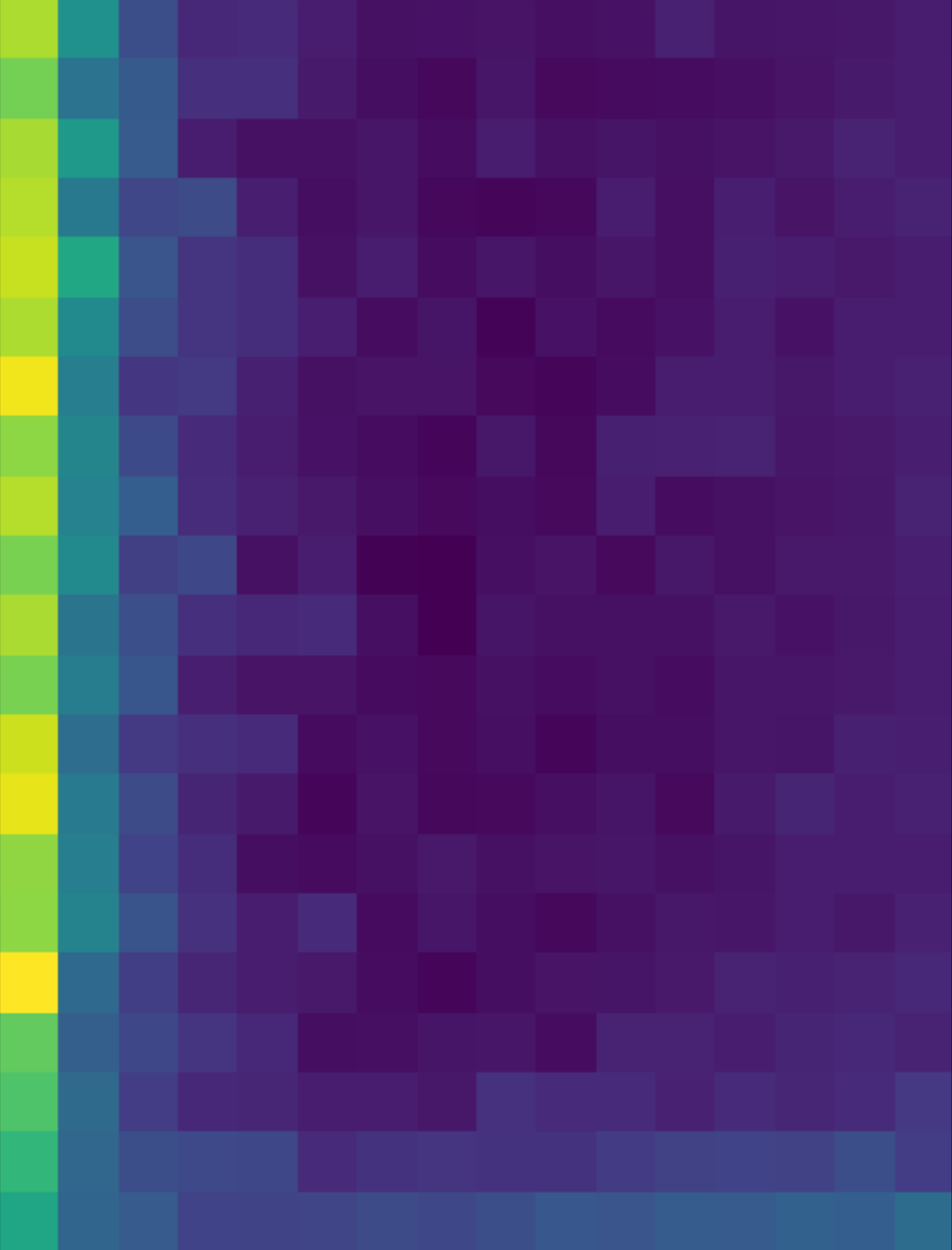};
\addplot [only marks, draw=none, mark=*, draw=red, fill=red, colormap/viridis]
table{%
x                      y
7 10
};
\end{axis}

\end{tikzpicture}
        \caption{2D heat map.}
        \label{fig:2d_heatmap}
    \end{subfigure}
    
    \caption{The results of UAM ecosystem design optimization through grid search under the Gaussian mixture demand model with the peak demand at 30 flight per hour. The optimal point is indicated with the red dot.}
    \label{fig:grid_search_optimization}
\end{figure}

\begin{figure}[t]
\centering
\small
\begin{subfigure}{.45\textwidth}
  \centering
\begin{tikzpicture}

\begin{groupplot}[group style={
    group size=1 by 2, 
    vertical sep=0.5cm,
    x descriptions at=edge bottom,
    y descriptions at=edge left,},
    xlabel={Fleet Size},]
\nextgroupplot[
legend cell align={left},
legend style={draw=white!80.0!black},
tick align=outside,
tick pos=left,
x grid style={white!69.01960784313725!black},
xmin=12.75, xmax=62.25,
xtick style={color=black},
y grid style={white!69.01960784313725!black},
ylabel={\(\displaystyle \Delta\) Optimal Cost},
ymin=-8.29697207561729, ymax=138.521347924383,
ytick style={color=black},
width=2.6in,
height=2.2in,
]

\addlegendimage{semithick, black, dashed}\addlegendentry{Vicinity bound}

\addplot [only marks, draw=none, mark=*, draw=red, fill=red, colormap/viridis]
table{%
x                      y
36 0
};
\addlegendentry{Optimal point}
\path [draw=blue, semithick]
(axis cs:15,120.795587924383)
--(axis cs:15,131.847787924383);

\path [draw=blue, semithick]
(axis cs:18,59.2683879243827)
--(axis cs:18,70.0827879243827);

\path [draw=blue, semithick]
(axis cs:21,27.8090879243827)
--(axis cs:21,39.3270879243827);

\path [draw=blue, semithick]
(axis cs:24,9.52448792438271)
--(axis cs:24,25.3162879243827);

\path [draw=blue, semithick]
(axis cs:27,4.95358792438271)
--(axis cs:27,11.4139879243827);

\path [draw=blue, semithick]
(axis cs:30,0.449187924382717)
--(axis cs:30,6.81158792438272);

\path [draw=blue, semithick]
(axis cs:33,-1.53711207561729)
--(axis cs:33,4.10248792438271);

\path [draw=blue, semithick]
(axis cs:36,-1.62341207561728)
--(axis cs:36,1.62338792438272);

\path [draw=blue, semithick]
(axis cs:39,0.584187924382712)
--(axis cs:39,2.91858792438271);

\path [draw=blue, semithick]
(axis cs:42,1.06598792438272)
--(axis cs:42,4.72298792438272);

\path [draw=blue, semithick]
(axis cs:45,3.67438792438272)
--(axis cs:45,6.57918792438272);

\path [draw=blue, semithick]
(axis cs:48,5.60798792438271)
--(axis cs:48,5.80198792438271);

\path [draw=blue, semithick]
(axis cs:51,8.28048792438272)
--(axis cs:51,8.98628792438272);

\path [draw=blue, semithick]
(axis cs:54,10.2710879243827)
--(axis cs:54,10.9674879243827);

\path [draw=blue, semithick]
(axis cs:57,13.2393879243827)
--(axis cs:57,13.9797879243827);

\path [draw=blue, semithick]
(axis cs:60,15.5759879243827)
--(axis cs:60,16.4079879243827);

\addplot [semithick, blue, mark=-, mark size=4, mark options={solid}, only marks]
table {%
15 120.795587924383
18 59.2683879243827
21 27.8090879243827
24 9.52448792438271
27 4.95358792438271
30 0.449187924382717
33 -1.53711207561729
36 -1.62341207561728
39 0.584187924382712
42 1.06598792438272
45 3.67438792438272
48 5.60798792438271
51 8.28048792438272
54 10.2710879243827
57 13.2393879243827
60 15.5759879243827
};
\addplot [semithick, blue, mark=-, mark size=4, mark options={solid}, only marks]
table {%
15 131.847787924383
18 70.0827879243827
21 39.3270879243827
24 25.3162879243827
27 11.4139879243827
30 6.81158792438272
33 4.10248792438271
36 1.62338792438272
39 2.91858792438271
42 4.72298792438272
45 6.57918792438272
48 5.80198792438271
51 8.98628792438272
54 10.9674879243827
57 13.9797879243827
60 16.4079879243827
};
\addplot [semithick, blue]
table {%
15 126.321687924383
18 64.6755879243827
21 33.5680879243827
24 17.4203879243827
27 8.18378792438271
30 3.63038792438272
33 1.28268792438271
36 -1.20756172847791e-05
39 1.75138792438271
42 2.89448792438272
45 5.12678792438272
48 5.70498792438271
51 8.63338792438272
54 10.6192879243827
57 13.6095879243827
60 15.9919879243827
};
\addlegendentry{$\Delta$ Optimal cost}
\addplot [semithick, black, dashed]
table{%
x  y
30 -20
30 200
};
\addplot [semithick, black, dashed]
table{%
x  y
42 -20
42 200
};

\nextgroupplot[
legend cell align={left},
legend style={at={(0.97,0.03)}, anchor=south east, draw=white!80.0!black},
tick align=outside,
tick pos=left,
x grid style={white!69.01960784313725!black},
xmin=12.75, xmax=62.25,
xtick style={color=black},
y grid style={white!69.01960784313725!black},
ylabel={\(\displaystyle \Delta\) Vertiport Capacity},
ymin=-1.095, ymax=0.995,
ytick style={color=black},
width=2.6in,
height=2.2in,
]

\addplot [semithick, mark=diamond*, blue, mark size=2, mark options={solid}]
table {%
15 -1
18 -0.7
21 0.4
24 0.8
27 -0.4
30 -0.3
33 0.1
36 0
39 0.5
42 0.4
45 0.1
48 -0.3
51 0.9
54 0.5
57 0.2
60 -0.1
};
\addlegendentry{$\Delta$ Vertiport Capacity}

\addplot [only marks, mark=*, draw=red, fill=red, colormap/viridis]
table{%
x                      y
36 0
};

\addplot [semithick, black, dashed]
table{%
x  y
30 -2
30 2
};
\addplot [semithick, black, dashed]
table{%
x  y
42 -2
42 2
};

\end{groupplot}

\end{tikzpicture}
  \caption{Changes of the optimal cost and the corresponding optimal normalized vertiport capacity as the fleet size varies from its optimal point.}
  \label{fig:sensitivity_fleetsize}
\end{subfigure}%
\hspace{0.6cm}
\begin{subfigure}{.45\textwidth}
  \centering
\begin{tikzpicture}

\begin{groupplot}[group style={
    group size=1 by 2, 
    vertical sep=0.5cm,
    x descriptions at=edge bottom,
    y descriptions at=edge left,},
    xlabel={Normalized Vertiport Capacity},]
\nextgroupplot[
legend cell align={left},
legend style={draw=white!80.0!black},
tick align=outside,
tick pos=left,
x grid style={white!69.01960784313725!black},
xmin=-0.1, xmax=2.1,
xtick style={color=black},
y grid style={white!69.01960784313725!black},
ylabel={\(\displaystyle \Delta\) Optimal Cost},
ymin=-4.27944207561729, ymax=52.8816179243827,
ytick style={color=black},
width=2.6in,
height=2.2in,
]

\addlegendimage{semithick, black, dashed}\addlegendentry{Vicinity bound}

\addplot [only marks, draw=none, mark=*, draw=red, fill=red, colormap/viridis]
table{%
x                      y
1 0
};
\addlegendentry{Optimal point}
\path [draw=blue, semithick]
(axis cs:0,35.3177879243827)
--(axis cs:0,50.2833879243827);

\path [draw=blue, semithick]
(axis cs:0.1,21.4866879243827)
--(axis cs:0.1,31.7940879243827);

\path [draw=blue, semithick]
(axis cs:0.2,9.27138792438272)
--(axis cs:0.2,19.1565879243827);

\path [draw=blue, semithick]
(axis cs:0.3,6.36948792438272)
--(axis cs:0.3,8.59888792438272);

\path [draw=blue, semithick]
(axis cs:0.4,2.35878792438271)
--(axis cs:0.4,5.41298792438271);

\path [draw=blue, semithick]
(axis cs:0.5,2.79338792438272)
--(axis cs:0.5,6.17238792438272);

\path [draw=blue, semithick]
(axis cs:0.6,3.98558792438271)
--(axis cs:0.6,8.27918792438271);

\path [draw=blue, semithick]
(axis cs:0.7,0.449187924382717)
--(axis cs:0.7,6.81158792438272);

\path [draw=blue, semithick]
(axis cs:0.8,1.52358792438271)
--(axis cs:0.8,5.82738792438271);

\path [draw=blue, semithick]
(axis cs:0.9,3.28008792438272)
--(axis cs:0.9,8.30908792438272);

\path [draw=blue, semithick]
(axis cs:1,-1.62341207561728)
--(axis cs:1,1.62338792438272);

\path [draw=blue, semithick]
(axis cs:1.1,-1.68121207561729)
--(axis cs:1.1,2.16358792438271);

\path [draw=blue, semithick]
(axis cs:1.2,1.74568792438272)
--(axis cs:1.2,8.31228792438272);

\path [draw=blue, semithick]
(axis cs:1.3,0.756287924382715)
--(axis cs:1.3,6.91148792438272);

\path [draw=blue, semithick]
(axis cs:1.4,1.06598792438272)
--(axis cs:1.4,4.72298792438272);

\path [draw=blue, semithick]
(axis cs:1.5,0.584187924382712)
--(axis cs:1.5,2.91858792438271);

\path [draw=blue, semithick]
(axis cs:1.6,3.02788792438272)
--(axis cs:1.6,10.9602879243827);

\path [draw=blue, semithick]
(axis cs:1.7,1.05008792438272)
--(axis cs:1.7,5.24888792438272);

\path [draw=blue, semithick]
(axis cs:1.8,4.62378792438272)
--(axis cs:1.8,9.50378792438272);

\path [draw=blue, semithick]
(axis cs:1.9,1.21748792438272)
--(axis cs:1.9,7.25348792438272);

\path [draw=blue, semithick]
(axis cs:2,5.32178792438272)
--(axis cs:2,12.8195879243827);

\addplot [semithick, blue, mark=-, mark size=4, mark options={solid}, only marks, legend={}]
table {%
0 35.3177879243827
0.1 21.4866879243827
0.2 9.27138792438272
0.3 6.36948792438272
0.4 2.35878792438271
0.5 2.79338792438272
0.6 3.98558792438271
0.7 0.449187924382717
0.8 1.52358792438271
0.9 3.28008792438272
1 -1.62341207561728
1.1 -1.68121207561729
1.2 1.74568792438272
1.3 0.756287924382715
1.4 1.06598792438272
1.5 0.584187924382712
1.6 3.02788792438272
1.7 1.05008792438272
1.8 4.62378792438272
1.9 1.21748792438272
2 5.32178792438272
};
\addplot [semithick, blue, mark=-, mark size=4, mark options={solid}, only marks, legend={}]
table {%
0 50.2833879243827
0.1 31.7940879243827
0.2 19.1565879243827
0.3 8.59888792438272
0.4 5.41298792438271
0.5 6.17238792438272
0.6 8.27918792438271
0.7 6.81158792438272
0.8 5.82738792438271
0.9 8.30908792438272
1 1.62338792438272
1.1 2.16358792438271
1.2 8.31228792438272
1.3 6.91148792438272
1.4 4.72298792438272
1.5 2.91858792438271
1.6 10.9602879243827
1.7 5.24888792438272
1.8 9.50378792438272
1.9 7.25348792438272
2 12.8195879243827
};
\addplot [semithick, blue]
table {%
0 42.8005879243827
0.1 26.6403879243827
0.2 14.2139879243827
0.3 7.48418792438272
0.4 3.88588792438271
0.5 4.48288792438272
0.6 6.13238792438271
0.7 3.63038792438272
0.8 3.67548792438271
0.9 5.79458792438272
1 -1.20756172847791e-05
1.1 0.241187924382714
1.2 5.02898792438272
1.3 3.83388792438272
1.4 2.89448792438272
1.5 1.75138792438271
1.6 6.99408792438272
1.7 3.14948792438272
1.8 7.06378792438272
1.9 4.23548792438272
2 9.07068792438272
};
\addlegendentry{$\Delta$ Optimal cost}

\addplot [semithick, black, dashed]
table{%
x  y
0.8 -20
0.8 80
};
\addplot [semithick, black, dashed]
table{%
x  y
1.2 -20
1.2 80
};

\nextgroupplot[
legend cell align={left},
legend style={at={(0.97,0.03)}, anchor=south east, draw=white!80.0!black},
tick align=outside,
tick pos=left,
x grid style={white!69.01960784313725!black},
xmin=-0.1, xmax=2.1,
xtick style={color=black},
xtick={0,0.5,1,1.5,2},
xticklabels={1.0, 1.5, 2.0, 2.5, 3.0},
y grid style={white!69.01960784313725!black},
ylabel={\(\displaystyle \Delta\) Fleet Size},
ymin=-12.9, ymax=6.9,
ytick style={color=black},
width=2.6in,
height=2.2in,
]

\addplot [semithick, mark=diamond*, blue, mark size=2, mark options={solid}]
table {%
0 -12
0.1 -6
0.2 0
0.3 6
0.4 0
0.5 6
0.6 -6
0.7 -6
0.8 1
0.9 0
1 0
1.1 0
1.2 0
1.3 0
1.4 6
1.5 3
1.6 0
1.7 3
1.8 0
1.9 0
2 6
};
\addlegendentry{$\Delta$ Fleet Size}

\addplot [only marks, draw=none, mark=*, draw=red, fill=red, colormap/viridis]
table{%
x                      y
1 0
};

\addplot [semithick, black, dashed]
table{%
x  y
0.8 -20
0.8 20
};
\addplot [semithick, black, dashed]
table{%
x  y
1.2 -20
1.2 20
};

\end{groupplot}

\end{tikzpicture}
  \caption{Changes of the optimal cost and the corresponding optimal fleet size as the vertiport capacity varies from its optimal point.}
  \label{fig:sensitivity_redundancy}
\end{subfigure}
\caption{The sensitivity of the optimal point.}
\label{fig:optimal_point_sensitivity}
\end{figure}
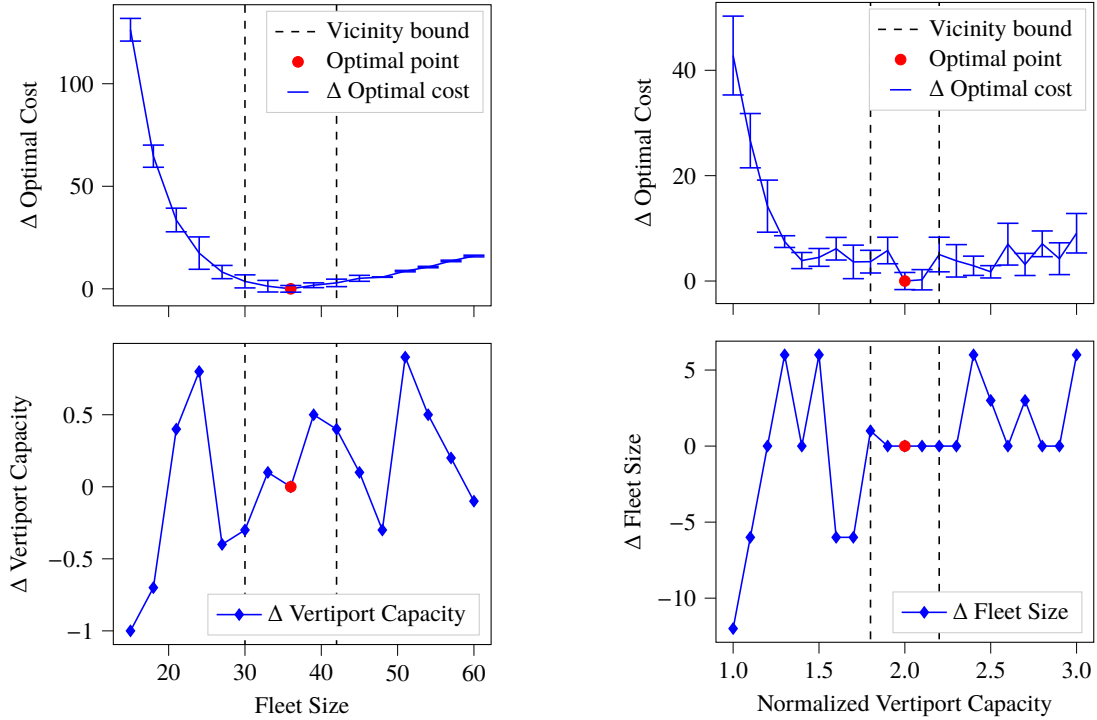

The results of UAM ecosystem design optimization through grid search are plotted in \cref{fig:grid_search_optimization} where \cref{fig:3d_surface} shows a 3D surface plot for the cost, and \cref{fig:2d_heatmap} shows a 2D heat map for the cost. 
The optimal design point through grid search is a fleet size of 36 and a normalized vertiport capacity of 2.0, with the optimal cost being 35.9.
We can see from the landscape of the cost function that it is high under small fleet size and low vertiport capacity, where it is more sensitive to the fleet size design variable. 
The cost is relatively flat as fleet size and vertiport capacity exceeds sensible threshold levels. 
This relatively flat landscape in the cost function indicates that once reasonable values have been chosen for long-term strategic and short-term strategic design variables like vertiport capacity and fleet size, the performance of the UAM ecosystem is most impacted by fleet management policy as highlighted in the previous section. 

We explore the sensitivities of the cost to each design variable further. 
\cref{fig:optimal_point_sensitivity} analyzes the sensitivity of each variable at the optimal point. 
We vary one of the design parameters while treating the other as a fixed value and evaluating the deviations from the optimal cost upon the variations of a single design variable.
The results of sensitivity analysis shed light on how designers of a UAM ecosystem can evaluate the risk of a potential wrong decision about a design parameter. 
A design parameter that leads to less sensitivity in the cost may need minimal adjustment or none at all to improve the performance of the UAM ecosystem. 
\cref{fig:sensitivity_fleetsize} shows the sensitivity of the optimal cost and vertiport capacity with respect to fleet size. 
Optimal cost is shown to have minimal sensitivity to changes in fleet size around its optimal value. 
We see a similar trend for changes in vertiport capacity, as the relatively large fluctuation in fleet size does not lead to significant fluctuation in the cost. 
This indicates that the cost, and thus ecosystem performance has low sensitivity for vertiport capacity values near the optimal point.
\cref{fig:sensitivity_redundancy} shows the sensitivity of the optimal cost and fleet size with respect to normalized vertiport capacity.
The sensitivities to changes in the fleet size and vertiport capacity design variables are again minimal. 
Both results imply that vertiport capacity and fleet size are low risk variables in the UAM ecosystem design process for the configurations tested in this work. 
Once a sensible value for these variables has been picked, the ecosystem is likely to be high performing even if the capacity value is not the optimal for the configuration in question. 
We again note, that the performance of the ecosystem is tightly coupled to the fleet management policies, and the results shown here are valid when proper fleet re-balancing and management policies are applied. 


\section{Conclusion and Future Work}

In this work, we examined the performance of a UAM ecosystem operating under nominal conditions using a variety of UAM ecosystem design choices in simulation.
We categorized these design variables into categories that represented decisions on the strategic long-term, strategic short-term, and tactical time horizons.
By evaluating the performance metrics such as throughput and delay time in the ecosystem, we were able to quantify the impact of various design choices on the operational ecosystem.
We found that design variables that fall into the tactical time horizon category such as choice of fleet and traffic management policies have a significant impact on system performance.
While design variable corresponding to long-term and short-term strategic decisions such as vertiport capacity and fleet size were found to have less impact on system performance so long as they fell into a suitable range of values.
The strategic design variables are generally considered difficult to modify and typically represent design choices that must be made months or even years prior to seeing the design choice operational.  
In short, we show that it is possible to design a performant UAM ecosystem, even when the vertiport network and the fleet serving it have not been optimized for the emergent demand profile.
Instead, an effective fleet management policy can play a significant role in how the system performs operationally, and must be optimized to . 

For future work, we plan to explore how our results generalize to more complex system disturbances and off-nominal operational profiles.
Specifically, we want to examine whether fleet and traffic management play as critical of a role under these conditions as they do during nominal operations. 
A wider range of design variables will also be explored in future work such as more complex vertiport configurations.
Additionally, we plan to explore more complex fleet management policies, and derive their theoretical performance properties in the UAM ecosystem. 
Lastly, we would like to consider how multiple operators that share resources in the UAM ecosystem impact the total system performance.
The design choices in a multi-operator UAM ecosystem would be driven by both the system managing coordination between operators, such as a UTM, the resource allocation policies for resources that are shared, and the fleet management policies each operator chooses to implement. 
Such design choices have been shown to have significant impact on the efficiency and the fairness of the ecosystem as a whole~\cite{evans2020fairness, chin2020tradeofffs}, and must be better understood in the context of the broader UAM ecosystem. 

\bibliography{references}

\end{document}